\documentclass[a4paper,reprint,aps,pra,showpacs,supscriptaddress,amsmath,amsfonts,amssymb]{revtex4-1}

\pdfoutput=1

\usepackage{graphicx}
\usepackage{nicefrac}
\usepackage[colorlinks,citecolor=blue,linkcolor=blue]{hyperref}
\usepackage{color}

\newcommand{\tj}[6]{ \left(\!\!\begin{array}{ccc}
  #1 & \!\!#2 & \!\!#3 \\
  #4 & \!\!#5 & \!\!#6
\end{array}\!\!\right)}
\newcommand{\Gj}[6]{ \left\{\!\!\begin{array}{ccc}
  #1 & \!\!#2 & \!\!#3 \\
  #4 & \!\!#5 & \!\!#6
\end{array}\!\right\}}

\begin{document}
\title{Quantum Noise for Faraday Light Matter Interfaces}

\author{D. V. Vasilyev}
\affiliation{Institute for Theoretical Physics and Institute for Gravitational Physics, Leibniz Universit\"at Hannover, Callinstr. 38, 30167 Hannover, Germany}
\author{K. Hammerer}
\affiliation{Institute for Theoretical Physics and Institute for Gravitational Physics, Leibniz Universit\"at Hannover, Callinstr. 38, 30167 Hannover, Germany}

\author{N. Korolev}
\affiliation{Niels Bohr Institute, Danish Research Foundation Center for Quantum Optics (QUANTOP), Blegdamsvej, DK-2100 Copenhagen, Denmark}
\author{A. S. S{\o}rensen}
\affiliation{Niels Bohr Institute, Danish Research Foundation Center for Quantum Optics (QUANTOP), Blegdamsvej, DK-2100 Copenhagen, Denmark}
\pacs{03.67.-a, 42.50.Lc, 42.50.Ct, 42.50.Ex}

\begin{abstract}
In light matter interfaces based on the Faraday effect quite a number of quantum information protocols have been successfully demonstrated. In order to further increase the performance and fidelities achieved in these protocols a deeper understanding of the relevant noise and decoherence processes needs to be gained. In this article we provide for the first time a complete description of the decoherence from spontaneous emission. We derive from first principles the effects of photons being spontaneously emitted into unobserved modes. Our results relate the resulting decay and noise terms in effective equations of motion for collective atomic spins and the forward propagating light modes to the full atomic level structure. We illustrate and apply our results to the case of a quantum memory protocol. Our results can be applied to any 
Alkali atoms, and the general approach taken in this article can be applied to light matter interfaces and quantum memories based on different mechanisms.
\end{abstract}

\maketitle

\section{Introduction}
The strong and coherent interaction of light with matter is a prerequisite for many approaches towards quantum information technologies. In particular long distance quantum communication relies on efficient light-matter interfaces which allow for a coherent transfer of quantum information from light to stationary carriers and back \cite{Kimble:2008if}. Also architectures for quantum computations based on light will depend on efficient light-matter interfaces for buffering and storing quantum information carried by light.

Optically dense atomic ensembles have proven to be a particularly promising technique for achieving strong coherent light matter interaction. Thereby quantum states of propagating pulses of light are mapped onto states of collective atomic (pseudo) spins. This essentially requires a mechanism to fiducially and reversibly convert photons into ground state spin excitations with long coherence times. Several mechanisms have been explored for realizing such an atomic ensemble based light matter interface, e.g. Raman transitions, Electromagnetically Induced Transparency \cite{Gorshkov:2007er,Gorshkov:2007gd,Gorshkov:2007kz}, Spin Echoes \cite{Tittel2009}, and Faraday rotation, see \cite{Andersreview,Sangouard11,Lvovsky:2009fr} for comprehensive reviews. The Faraday effect --- which will be the topic of this article --- consists in the rotation of light polarization depending on atomic spin polarization, and vice versa. In the context of light matter interface it has been successfully used to create squeezed states for spin \cite{Kuzmich:2000hv,Takano:2009bl} and light \cite{Mikhailov2008,Agha2010}, entangled states of collective atomic ensembles \cite{Julsgaard:2001by,Krauter:2011fj}, quantum teleportation from states of light to atoms \cite{Sherson:2006dt}, and quantum memory for light \cite{Polzik,Jensen:2010ba}. Apart from the light-matter interface the Faraday interaction has important applications also in continuous nondemolition measurement of atomic spin ensembles \cite{Smith2004,Chaudhury2006}, quantum-state control/tomography \cite{Deutsch:2010}, and magnetometry \cite{Wasilewski:2010cl,Napolitano2010,Napolitano:2011}.

These protocols can all be understood in terms of a rather simple model of the Faraday interaction. In this model atoms are assumed to have a spin $\nicefrac12$ ground state and a spin $\nicefrac12$ excited state \cite{Duan:2000ip}. Far off resonant light probing this dipole allowed transition will then experience a polarization rotation due to the dipole selection rules for such a  $\nicefrac12\rightarrow\nicefrac12$ transition. By the same effect the atomic polarization will be rotated by light. For far off resonant light atoms will be only very weakly excited such that this birefringence of the atomic medium can be understood as being due to the polarizability of the atomic ground states alone \cite{Happer:1967jv,Happer:1972fi}. The \textit{coherent} interaction of light with atoms arises from the \textit{real} part of the atomic polarizability. By the Kramers-Kronig relation it is clear that the corresponding imaginary part will necessarily be non-zero and add some \textit{incoherent} effects to the dynamics. Sure enough, these effects can be understood as resulting from spontaneous emission events. They will cause both, decoherence of light (absorption) and decoherence of atoms (spin decay). While both effects can be kept small as compared to the coherent dynamics, they are ultimately unavoidable on a fundamental level.

\begin{figure}[tbp] 
   \centering
\includegraphics[width=7cm]{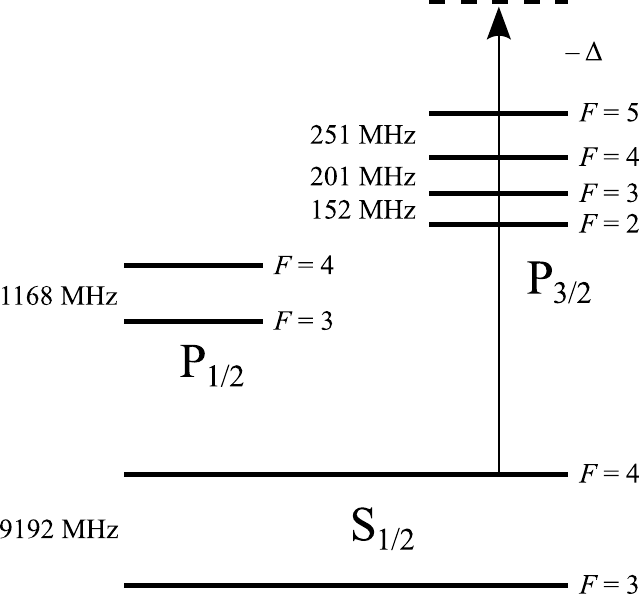}
\caption{Level structure of Cesium. We assume that our laser light is blue detuned by an amount $-\Delta$ from the transition $F=4\to{}F'=5$}
   \label{fig:level_scheme}
\end{figure}

Both decoherence effects --- light losses and spin decay --- can be included on the level of the simple spin $\nicefrac12$ model giving a qualitative understanding of the tradeoff between coherent and incoherent contributions to the light matter interaction \cite{Hammerer:2004je,Madsen:2004uq,Koschorreck2009}. However, in view of the experimental achievements it will become increasingly important to further gain a detailed and more quantitative understanding of these effects and the resulting tradeoff. A theoretical description based on a realistic atomic level scheme, such as the one shown in Fig.~\ref{fig:level_scheme}, and starting from first principles has not been given so far. In this article we will provide such a description.

From the standard dipole interaction of a single multi-level atom with the three dimensional electromagnetic field we consistently derive effective equations of motion for the collective ground state spin and the forward-propagating light modes relevant to the description of the light matter interface. In the derivation we will keep track of the effects of events where photons are emitted to any other than the forward direction. This eventually adds decay and noise terms to the equations of motions whose origin and dependence on the details of the atomic level structure are fully understood and explained for the first time in the present article. Important steps towards a deeper understanding have been taken before in \cite{Martin}. The insight gained by the detailed knowledge of losses and decoherence will enable an optimized operation of the light matter interface based on the Faraday interaction, and therefore contribute to boost its performance and fidelity.

The article is organized as follows: In section~\ref{sec:interaction} we will briefly summarize the Faraday rotation effect and connect it to the atomic polarizability. In section~\ref{sec:coherent_interaction} we give a description of the coherent part of the dynamics, deriving effective equations of motion for collective atomic spins and forward propagating modes. We will write down and solve these equations of motion also for canonical operators for light and atoms, and illustrate the working principle of a quantum memory protocol. In section~\ref{sec:decoherence} we arrive at the main results of this article and include in our derivation spontaneous emission and decoherence starting from first principles. The resulting equations of motion can again be expressed and integrated in terms of canonical operators. As an illustration we apply the result to the quantum memory protocol. In section~\ref{sec:summary} we provide a self-contained executive summary of our results. Readers just interested in applying the correct model for decoherence in the Faraday based light matter interface can directly consult this section.


\section{Effective interaction}\label{sec:interaction}

We consider alkali atoms, which have a single electron outside a closed shell. This electron will be in a $S_{1/2}$-state, giving two stable ground levels: a higher - and a lower hyperfine manifold, describing respectively a state where the electronic spin $J=\frac{1}{2}$ (in this article we put $\hbar=c=1$) is parallel to the nuclear spin $I$ giving a total of $F=I+\frac{1}{2}$ and a state where the spins are pointing opposite giving $F=I-\frac{1}{2}$, see Fig.~\ref{fig:level_scheme} where we show the level structure of the specific case of Cs where $I=\frac72$ so that $F=3,4$. There will be two dipole-allowed transitions to the excited states $P_{1/2}$ and $P_{3/2}$, each of which consists of several hyperfine states with spin $F'$ (primed variables always refer to electronically excited states). In this article we will mainly consider the case where atoms are initially prepared in the $S_{1/2}(F=I+\frac{1}{2})$-state and are driven on the $D_2$-transition to $P_{3/2}$ as indicated in Fig.~\ref{fig:level_scheme} for the example of $^{133}$Cs.

In the standard dipole and rotating wave approximation the coupling of light to atoms is given by
\begin{align*}
H_{\rm{int}}&=
-(\mathbf{d}^{(+)}\mathbf{E}^{(-)}+\mathbf{d}^{(-)}\mathbf{E}^{(+)}).
\end{align*}
where $\mathbf{E}^{(\pm)}$ and $\mathbf{d}^{(\pm)}$ are the positive/negative frequency component of, respectively, the electric field $\mathbf{E}=\mathbf{E}^++\mathbf{E}^-$ and the dipole moment operator $\mathbf{d}=\mathbf{d}^++\mathbf{d}^-$. We will be interested here in the interaction with far off-resonant light only. In this case the excited states will be only very weakly populated. In the limit of low saturation excited states can be adiabatically eliminated (Appendix~\ref{app:adiabat-elim}), and the light-atom interaction is described by the effective Hamiltonian
	\begin{equation}\label{eff_int}
	H_{\rm{int}}^{\rm{eff}} = \mathbf{E}^{(-)}{\boldsymbol\alpha}\,\mathbf{E}^{(+)}.
	\end{equation}
We have here introduced the polarizability tensor operator
	\begin{align}\label{pol_tens}
	\boldsymbol\alpha&=-\sum_{F'}\frac{P_g\mathbf{d}P_{F'}\mathbf{d}P_g}{\Delta_{F'}},
	\end{align}
where the projection operators are defined as
	\begin{align}
	P_F &=\sum_m|F,m{\rangle}{\langle}F,m|,
	&P_g &=\sum_FP_F,\\
	P_{F'}&=\sum_{m'}|F'\!,m'{\rangle}{\langle}F'\!,m'|, 
	&P_e &=\sum_{F'}P_{F'},
	\end{align}
such that $P_g+P_e=1$. Here $\Delta_{F'}$ is the detuning of the light from the $S_{1/2}F\rightarrow P_{3/2}F'$ transition, see Fig.~\ref{fig:level_scheme}. The polarizability operator $\boldsymbol\alpha$ in \eqref{pol_tens} has to be understood as a $3\times 3$ matrix whose Cartesian components are given by $\boldsymbol\alpha_{ij}=-\sum_{F'} P_g\mathbf{d}_iP_{F'}\mathbf{d}_jP_g/\Delta_{F'}$. Each of these components is an operator acting exclusively within the subspace of ground states $S_{1/2}(F=I\pm\frac{1}{2})$.

The effective interaction in \eqref{eff_int} describes second order processes where a photon is absorbed and reemitted, while the atom makes a transition from a ground to an excited state and back again to a ground state. In such a process the initial and final ground states can either have the same or different total spin $F=I\pm\frac{1}{2}$. In order to describe processes involving transitions within the $F$-subspace we introduce $H_{FF}=P_FH_{\rm{int}}^{\rm{eff}}P_F$. Transitions between $F\leftrightarrow F-1$ we describe by $H_{FF-1}=P_{F-1}H_{\rm{int}}^{\rm{eff}}P_F+P_{F}H_{\rm{int}}^{\rm{eff}}P_{F-1}$.
In general such second order processes could also give rise to changes of $F$ by 2, but since we are considering $S_{1/2}$ ground levels there are only two ground levels, \textit{e.g.} as shown in Fig.~\ref{fig:level_scheme} for Cs $F=3,4$.

\begin{figure}[tbp]
\centering
\includegraphics[width=8.5cm]{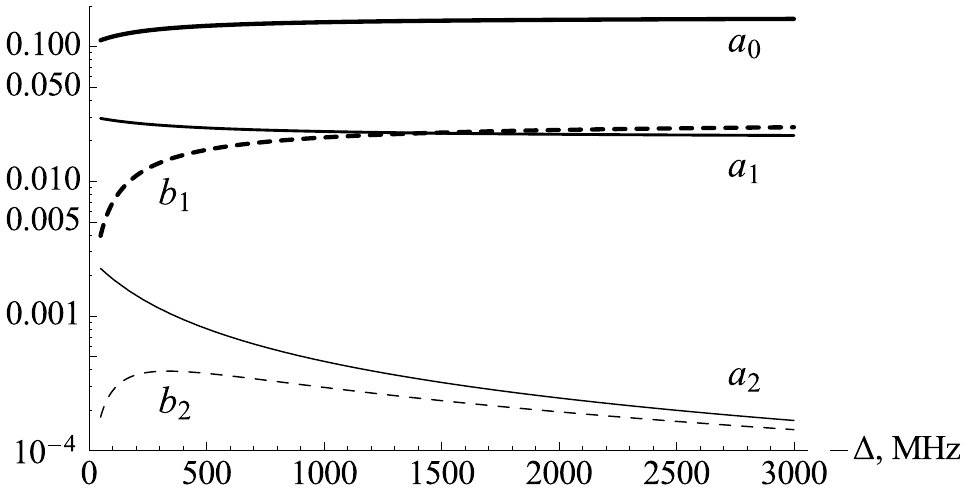}
\caption{$a$ and $b$ coefficients as functions of detuning $-\Delta$ in MHz for Cesium with $F=4$; $a_0$~(thick), $a_1$~(medium), $a_2$~(thin), $b_1$~(thick dashed), $b_2$~(thin dashed). In the limit of high detuning we obtain 
$a_{0}\to\frac16$, $a_{1}\to\frac1{48}$, $b_{1}\to\frac1{16\sqrt5}$, and $\{a_{2},\,b_{2}\}\to 0$.}
\label{fig:ab_coeff}
\end{figure}

Consider first transitions within one $F$-subspace as described by the Hamiltonian $H_{FF}$. Due to the fact that we have conservation of angular momentum in the interaction, we can decompose the polarizability operator $\alpha_{FF}=P_F\boldsymbol\alpha P_F$ into its tensor parts \cite{Happer:1972fi,Deutsch:2010}, namely
	\begin{equation}\label{pol_decomp}
	\alpha_{FF} = -\frac{d_0^2}{\Delta}(a_0 + ia_1\mathbf{j}\times+a_2 \mathbf{Q}).
	\end{equation}
Here the dipole matrix element is defined as $d_0^{2}=(2J'+1)|\langle J'\|d\|J\rangle|^{2}$ with $J$ and $J'$ being the electronic angular momenta of the ground and exited states, respectively (see Appendix~\ref{A:hamiltonian_constraction}), $\Delta$ is the detuning from resonance as shown in Fig.~\ref{fig:level_scheme}. The three terms are the scalar, vector and second rank tensor part, respectively. $\mathbf{j}$ is the spin operator for the total ground state spin, that is $\mathbf{j}^2|F,m\rangle=F(F+1)|F,m\rangle$. In the vector component we use here the short hand notation
	\begin{equation*}
	\mathbf{j}\times=\left(\!\!
	\begin{array}{ccc}
	0 & j_z & j_y \\
	j_z & 0 & -j_x \\
	-j_y & j_x & 0 \end{array}\!\!\right),
	\end{equation*}
which is equivalent to taking the cross product of $\mathbf{j}$ with the vector to the right and then the scalar product with the vector to the left. One can use vector product properties in order to transform the expression as follows $i\,\mathbf{E}^{(-)}\cdot[\mathbf{j}\times\mathbf{E}^{(+)}]=-i\,\mathbf{j}\cdot[\mathbf{E}^{(-)}\times\mathbf{E}^{(+)}]$. The latter form could be more convenient since it is just a scalar product of the spin and a vector for light which resembles the Stokes operator characterizing the polarization of light, as will be discussed in more detail below. The second rank tensor part $\mathbf{Q}$ is defined component-wise as
	\begin{align*}
	Q_{ij}&=-(j_ij_j+j_jj_i)+\delta_{ij}\frac{2}{3}\mathbf{j}^2.
	\end{align*}
The tensor decomposition of $\boldsymbol\alpha$ is given in detail in Appendix~\ref{A:hamiltonian_constraction} where we also give the explicit expressions for the real coefficients $a_0,a_1$ and $a_2$ and their dependence on the detuning. In Fig.~\ref{fig:ab_coeff} we show these coefficients for the case of Cs. Note in particular that the second rank tensor polarizability vanishes for large detunings.

For the other case where the final spin state has a different total spin than the initial state, we have
	\begin{equation}
	\alpha_{FF\pm1} = -\frac{d_0^2}{\Delta}(b_1\mathbf{T}^{(1)}+b_2\mathbf{T}^{(2)}).
	\end{equation}
The coefficients $b_1$ and $b_2$ and the irreducible tensor operators $\mathbf{T}^{(1)}$ and $\mathbf{T}^{(2)}$ 
 are given in Appendix~\ref{A:hamiltonian_constraction}. We do not have a scalar tensor component here because in these processes the spin state is changed such that the corresponding Hamiltonian cannot have a component proportional to the identity operator \textit{i.e.} a scalar component. In Fig.~\ref{fig:ab_coeff} we show $b_1$ and $b_2$ as a function of the detuning. Note that $b_1$ and $b_2$ have the same scale as $a_1$ and $a_2$, respectively. 


The effective interaction Hamiltonian \eqref{eff_int} describes elastic Rayleigh and inelastic Raman scattering of light on atoms. It is a Hermitian operator and as such gives rise to a fully coherent evolution for the overall system comprised of the atom and the quantized electromagnetic field. The quantum coherent effects resulting from this effective interaction have been widely used in the field of quantum information, both theoretical and experimental \cite{Polzik}. Note however, that within the regime of its validity --- that is far below saturation of excited states --- the interaction Hamiltonian \eqref{eff_int} \textit{still} contains and correctly describes the effects of spontaneous emission on both atom and light: If light is treated as a reservoir and we trace over the random position of the atoms we will show that we can describe spontaneous emission which leads to decoherence 
effects within the ground state levels. Vice versa, the atom (or an ensemble of atoms) can provide an effective absorptive medium for light. In the following sections we will first treat the \textit{coherent interaction} and then include the \textit{decoherence effects} resulting from the effective interaction Hamiltonian \eqref{eff_int}.


\section{Coherent interaction}\label{sec:coherent_interaction}

\subsection{Interaction with a single atom}
\begin{figure}[tbp]
\centering
\includegraphics[width=8.5cm]{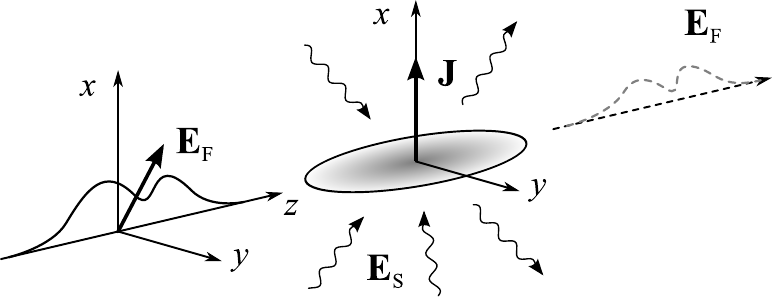}
\caption{The spin polarized atomic sample interacts with the off-resonant light field $\mathbf{E}_{\rm{F}}$ and non-forward modes $\mathbf{E}_{\rm S}$ in vacuum states.}
\label{fig:fig_scheme_setup}
\end{figure}
We consider now a situation as shown in Fig.~\ref{fig:fig_scheme_setup}. A cloud of atoms interacts with an incoming pulse of light described by the forward propagating electric field $\mathbf{E}_{\rm{F}}$, and we are interested in the coherent evolution of the spin of atoms and the forward propagating field modes \cite{Brian,Hammerer:thesis}. 
The total electric field $\mathbf{E}=\mathbf{E}_{\rm{F}}+\mathbf{E}_{\rm{S}}$ will be the sum of forward propagating modes and the remaining, non-forward propagating modes $\mathbf{E}_{\rm{S}}$, cf. Fig.~\ref{fig:fig_scheme_setup}. 
These modes will be treated as a reservoir in the later sections and give rise to the decoherence effects to be discussed there.

For the moment we will thus restrict ourselves to light propagating along the $z$-direction. In principle the forward modes will also experience some spreading \cite{Vasilyev08,Vasilyev11}, but here we take the opening angle to be roughly zero, and use a one dimensional description of $\mathbf{E}_{\rm{F}}(z)$. The separation into a forward propagating mode $\mathbf{E}_{\rm{F}}$ described by a one dimensional equation of motion and a set of non forward modes $\mathbf{E}_{\rm{S}}$ is further justified in Ref. \cite{Martin,Andersreview} and is valid if the Fresnel number of the ensemble is much larger than unity. In this model the field is described in terms of position dependent annihilation (creation) operators, defined by
	\begin{equation}
	a_\sigma(z,t) = \int\frac{dk}{2\pi}a_{k\sigma}(t)e^{ikz},
	\end{equation}
where $\sigma=x,y$ labels the transverse polarizations. These operators obey $[a_{\sigma}(z,t),a^{\dagger}_{\sigma'}(z',t)]=\delta_{\sigma\sigma'}\delta(z-z')$. In terms of $a_x(z,t)$ and $a_y(z,t)$ the Stokes operators $S_i(z,t)$ can be introduced
	\begin{align}
	\label{stockes_definition_begins}
	S_x&=\frac{1}{2}(a^{\dagger}_xa_x-a^{\dagger}_ya_y),&
	S_y&=\frac{1}{2}(a^{\dagger}_xa_y+a^{\dagger}_ya_x),\\
	S_z&=\frac{1}{2i}(a^{\dagger}_xa_y-a^{\dagger}_ya_x),&
	\label{stockes_definition_ends}
	S_0&=\frac{1}{2}(a^{\dagger}_xa_x+a^{\dagger}_ya_y),
	\end{align}
which are a convenient tool to describe the polarization state of light propagating along the $z$-direction. They obey the commutation relations for an angular momentum density $[S_i(z),S_j(z')]=i\delta(z-z')\epsilon_{ijk}S_k(z)$.

Each part in the decomposition \eqref{pol_decomp} of the atomic polarizability operator will give rise to a term in  the effective interaction Hamiltonian, such that $H_{FF}=H^{(0)}+H^{(1)}+H^{(2)}$ corresponding to the scalar, vector and tensor parts respectively. These terms can be conveniently expressed and interpreted in terms of Stokes operators. The scalar Hamiltonian --- the tensor-0 term is
	\begin{equation}
	H^{(0)} = g a_0 S_0.
	\end{equation}
Here we have defined $g=-\frac{d_0^2}{\Delta}\frac{\omega_0}{\epsilon_0A}$ where 
$\Delta$ is the detuning from resonance as shown in Fig.~\ref{fig:level_scheme}, $\omega_0$ the atomic transition frequency, and $A$ the beam cross section. It can be interpreted as a Stark shift, which equally shifts all atomic energy levels proportional to the light intensity. Conversely it can also be interpreted as an equal shift of the frequency for all light modes (off-resonant with a detuning $\Delta$), that is as a new index of refraction seen by light.

The vector Hamiltonian, which for our purpose is the most interesting part, is
	\begin{equation}
	H^{(1)} = ga_1S_zj_z.
	\end{equation}
$S_z$ describes the circularity of light and $j_z$ the $z$-component of the atomic spin. In the interaction the atomic spin is rotated around the $z$-axis, by an amount proportional to $S_z$. Likewise, the Stokes vector is rotated about $z$-axis, by an amount proportional to $j_z$. This is a circular birefringence effect and this interaction gives us the desired Faraday interaction.

Finally we also have a complicated tensor Hamiltonian
	\begin{equation}
	\label{H2_hamiltonian}
	H^{(2)}=-ga_2(S_{x}(j_x^2-j_y^2)+S_y\{j_x,j_y\}+2S_0(3j_z^2-\mathbf{j}^2)/3).
	\end{equation}
This amounts to a dynamical Stark shift. 
The effect vanishes for large detunings since the coefficient $a_{2}$ goes to zero as shown in Fig.~\ref{fig:ab_coeff}. It can be interesting for coherent dynamics in some experiments for tomography and collective squeezing purposes \cite{Wasilewski:2009en,Wasilewski:2010cl,Krauter:2011fj,Mishina2005,Mishina:2007kx}. We keep it for calculation of spontaneous emission, and neglect it for simplicity in the coherent dynamics.

\subsection{Interaction with an atomic ensemble}

So far we have considered a single atom. We will now assume that light is interacting with a cloud of $N_{a}$ atoms of length $L$ and constant atomic density $\rho$ as shown in Fig.~\ref{fig:fig_scheme_setup}. For the atomic spin we therefore introduce the continuous spin density operators, which are a sum of the single atom $(j^{a})$ angular momentum operators evaluated at their respective positions
	\begin{align}
	\label{theory:continuous_spin}
	j_k(z)&=\sum_{a=1}^{N_{a}}\delta(z-z_a)j_{k}^{a},\qquad k=x,y,z.
	\end{align}
Then the commutator is
	\begin{align}
	[j_m(z),j_n(z')]&=\sum_{a,b}^{N_{a}}\delta(z-z_a)\delta(z'-z_{b})[j_m^{a},j_n^{b}]\nonumber\\
	&=i\,\epsilon_{mnk}\,j_k(z)\delta(z-z')\label{spin_commutator_z},
	\end{align}
in perfect analogy to the commutation relation for the Stokes operators. For details regarding the definition of these continuous spin operators, in particular concerning transverse effects, we refer to \cite{Andersreview,Martin}.

We are now in a position to describe the coherent interaction of the forward propagating light modes with the atomic ensemble. We ignore here the contributions from $H^{(0)}$ since it merely accounts for the overall refractive index. The effective interaction can then be written
	\begin{equation}
	H^{\rm{eff}}_{\rm{coh}} = g\int_0^L(\boldsymbol{\gamma}\cdot\mathbf{S})(z,t)dz,
	\end{equation}
where $\boldsymbol{\gamma}=(-a_2(j_x^2-j_y^2),-a_2\{j_x,j_y\},a_1j_z)$
is a vector describing atomic polarization. In Appendix~\ref{A:light-transform} we show how the time derivative of the light operators can be transformed to a derivative in position ($z$), such that when we form the Heisenberg equation of motion we get from the relation $[S_i(z),S_j(z')]=i\delta(z-z')\epsilon_{ijk}S_k(z)$ that
\begin{equation}
\frac{\partial}{\partial{}z}\mathbf{S}(z,t) = g(\boldsymbol{\gamma}\times\mathbf{S})(z,t).
\end{equation}
If we replace the operators in $\boldsymbol{\gamma}$ with their
expectation values, we can see that in the interaction the Stokes operator $\mathbf{S}$ gets rotated about the vector $\boldsymbol{\gamma}$. Written out in full detail the equation reads
\begin{equation}
\label{stockes_big_equations}
\frac{\partial}{\partial{}z}\!{\left(\!\!\begin{array}{ccc}
S_x\\
S_y\\
S_z
\end{array}\!\!\right)}
\!\!=\! g\! \left(\!\!\begin{array}{ccc}
0 & -a_1j_z & \!\!\!\!\!-a_2\{j_x,j_y\} \\
a_1j_z & 0 & \!\!\!\!a_2(j_x^2-j_y^2) \\
a_2\{j_x,j_y\} & \!\!-a_2(j_x^2-j_y^2) & 0
\end{array}\!\!\right)
\!\!
\left(\!\!\begin{array}{ccc}
S_x\\
S_y\\
S_z
\end{array}\!\!\right)
\!\!.
\end{equation}
These equations have first been derived by B. Julsgaard \cite{Brian}. The rotation of $\mathbf{S}$ can be seen to be composed of a big
rotation (proportional to $a_1$) around the $z$-axis and proportional to
the atomic spin along $z$ and a small rotation (proportional to $a_2$)
in the $(x,y)$ plane by an angle which depends on the relative angle
between the mean atomic spin and the Stokes vector. 
If we only consider the $a_1$-terms above, and assume the $x$ component of Stokes vector to be much larger than the other two projections (as is the case for $x$ polarized light), and assuming the rotation angle to be small, then we arrive at the effectice equation of motion
	\begin{equation}
	\label{stockes_small_equations}
	\frac{\partial}{\partial{}z}\binom{S_y}{S_{z}}(z,t) = ga_1S_x \binom{j_z}{0}.
	\end{equation}
These simpler equations tell us that $S_z$ is conserved in the interaction, while $S_y$ receives some contribution from the spin component $j_{z}$.

A similar analysis can be performed for the the coherent dynamics of the atoms. We will write the coherent Hamiltonian as
\begin{equation}
H_{\rm{coh}}=g[\boldsymbol{\gamma}{\cdot}\mathbf{S}+{\gamma}_0S_0],
\end{equation}
where $\boldsymbol{\gamma}$ is given as above and $\gamma_0=-a_2j_z^2$. From this we can determine the coherent evolution of the spin vector $\mathbf{j}$
\begin{widetext}
\begin{equation}
\label{spin_big_equation}
\frac{\partial}{\partial{t}}{\left(\!\begin{array}{ccc}
j_x\\
j_y\\
j_z
\end{array}\!\!\right)}
=g\left[\left(\!\begin{array}{ccc}
a_2\{j_y,j_z\}& \!\!-a_2\{j_x,j_z\} & \!\!\!\!-a_1j_y \\
a_2\{j_x,j_z\} & a_2\{j_y,j_z\} & \!\!a_1j_x \\
\!\!-2a_2\{j_x,j_y\} & \!\!2a_2(j_x^2-j_y^2) & 0
\end{array}\!\!\right)
\left(\!\!\begin{array}{ccc}
S_x\\
S_y\\
S_z
\end{array}\!\!\right)
+{a_2}\left(\!\!\begin{array}{ccc}
-\{j_y,j_z\}\\
\{j_x,j_z\}\\
0
\end{array}\!\!\right)S_0\right].
\end{equation}
\end{widetext}
While the overall form of these equations is rather complicated we can get valuable insight by treating the different terms separately \cite{Brian}. For instance if we only consider $a_1$-terms as it was done with light and also assume small Faraday rotations of atomic spins with strong polarization along  $x$ then we get the equation
	\begin{equation}
	\label{spin_small_equation}
	\frac{\partial}{\partial{}t}\binom{j_y}{j_z}(z,t) = ga_1j_x\binom{S_z}{0},
	\end{equation}
in perfect analogy to the effective equations of motion for light found above, cf. \eqref{stockes_big_equations} and \eqref{stockes_small_equations}. $j_z$ is unaltered in the interaction, whilst $j_y$ gets a contribution from $S_z$.

\subsection{Canonical Operators}\label{subsec:Canonical-Operators}

In order to further emphasize the similarity in the atom and light evolution we will consider them on a more equal footing by introducing canonical variables. From the definition of the Stokes operators for light \eqref{stockes_definition_begins}--\eqref{stockes_definition_ends} one can see that for a large classical field polarized in the $x$-direction and sufficiently small angles of Faraday rotations the $S_{y}$ and $S_{z}$ components are proportional to the $X$ and $P$ quadratures of the 
weak quantum field in the orthogonal polarization ($y$).
	\begin{align}
	S_{y}(z,t) &= \sqrt{\langle S_{x}\rangle}X_{L}(z,t),\\
	S_{z}(z,t) &= \sqrt{\langle S_{x}\rangle}P_{L}(z,t).
	\end{align}
The quadratures introduced here obey standard commutation relation $[X_L(z,t),P_L(z',t)]=i\delta(z-z')$. 

Next we consider canonical spin densities. Experimentally two different configurations are typically employed. Either the light polarization is oriented along the same axis as the spins of the atoms ($x$) or the light is orthogonal to the atomic polarization ($y$). For simplicity we will restrict our discussion of the general method to the case where the light is polarized along the polarization of the spins. The argument we give can however easily be generalized also to the other orientations and for completeness we give the results for both orientations below. Let us assume that the atomic spin is polarized along the $x$-axis. Then the commutation relation \eqref{spin_commutator_z} averaged over the random positions of the atoms read
	\begin{equation}
	\overline{[j_{y}(z,t),j_{z}(z',t)]} = i\langle j_{x}\rangle\delta(z-z'),
	\end{equation}
where $\langle j_{x}\rangle  = n\langle j_{x}^{a}\rangle$ with $n$ being the average linear density of atoms. The field like canonical variables for the spin subsystem,
	\begin{equation}\label{atomic_canonical_definition}
	X_{A}(z,t) = \frac{j_{y}(z,t)}{\sqrt{ \langle j_{x}\rangle}},\qquad
	P_{A}(z,t) = \frac{j_{z}(z,t)}{\sqrt{ \langle j_{x}\rangle}},
	\end{equation}
obey the canonical commutation relation
	\begin{equation}
	\overline{[X_{A}(z,t),P_{A}(z',t)]} = i\delta(z-z').
	\end{equation}
Now we can write the equations of motion for light and atoms given by \eqref{stockes_small_equations},~\eqref{spin_small_equation} in terms of the canonical operators. Upon introducing the coupling constant $\kappa=ga_1\sqrt{\frac{F}{2}N_aN_p}$ where $N_{p}$ is a number of photons in the driving field one obtains
	\begin{equation}\label{EOM_canonical_small}
	\begin{split}
	\frac{\partial}{\partial{}z}\binom{X_L}{P_L}(z,t) &= \frac{\kappa}{\sqrt{LT}}\binom{P_{A}}{0},\\
	\frac{\partial}{\partial{}t}\binom{X_{A}}{P_{A}}(z,t) &= \frac{\kappa}{\sqrt{LT}}\binom{P_{L}}{0}.
	\end{split}
	\end{equation}
As opposed to the initial set of equations \eqref{stockes_big_equations}, \eqref{spin_big_equation} these simplified equations are not coupled to each other, since we have dropped the term $H^{(2)}$ of the interaction Hamiltonian (\ref{H2_hamiltonian}) due to the second rank tensor polarizability. This is justified in the limit of large detuning. Their solution can be expressed in the form of input-output relations for collective variables which are introduced by
	\begin{align}
	X_{L}^{in(out)} &= \frac1{\sqrt{T}}\int_{T} dt\, X_{L}(0(L),t),\\
	X_{A}^{in(out)} &= \frac1{\sqrt{L}}\int_{L} dz\, X_{A}(z,0(T)).
	\end{align}
Here $T$ is the duration of the light pulse and $L$ is the length of the atomic sample. The same definitions are applied to the conjugated canonical variables of light and atoms. Finally, the input-output relations obtained from the equations of motion \eqref{EOM_canonical_small} read
	\begin{align}
	\label{st:inoutL}
	\binom{X^{out}_L}{P^{out}_L} &= \binom{X^{in}_L}{P^{in}_L} + \kappa\binom{P^{in}_A}{0},\\
	\label{st:inoutA}
	\binom{X^{out}_A}{P^{out}_A} &= \binom{X^{in}_A}{P^{in}_A} + \kappa\binom{P^{in}_L}{0}.
	\end{align}
These relations constitute the foundation for the light--matter interface based on the Faraday rotation, see Ref. \cite{Andersreview}.

A specific example for the application of the Faraday interaction is the possibility to create a memory for light. The quantum state of a propagating pulse of light is thereby mapped onto the collective spin of a cloud of atoms. The mapping protocol applied in Ref. \cite{Polzik} works as follows: The light pulse first interacts with the atomic cloud such that the evolution is described by \eqref{st:inoutL} and \eqref{st:inoutA}. One then measures $X_L^{out}$ via homodyne detection and uses the result for a feedback on the atomic spin. The feedback should substract the measurement outcome for $X_L^{out}$ from $P_A^{out}$ with a gain $\nu$. One can show that the evolution  of the atoms is then described by the relations \cite{Polzik}
\begin{align}
X_A^{out}&=X_A^{in}+\kappa{P_L^{in}},\\
{P'}_{A}^{out}&=P_{A}^{out}-\nu X_{L}^{out}=P_{A}^{in}(1-\kappa \nu)-\nu X_{L}^{in}.
\end{align}
Assuming $\kappa=\nu=1$ we have
\begin{align}
X_A^{out}&=X_A^{in}+{P_L^{in}},\\
P_A^{out}&=-X_L^{in},
\end{align}
meaning that we have stored the light quadratures in atoms
\begin{align}
\langle{}X_A^{out}{\rangle}&={\langle}P_L^{in}{\rangle},\\
\langle{}P_A^{out}{\rangle}&=-{\langle}X_L^{in}{\rangle}.
\end{align}
 The quality of the mapping can be characterized by the fidelity $\mathcal{F}$.  
Assuming a random set of coherent states as the input state the fidelity is found to be \cite{Giedke:thesis}
\begin{equation}
\label{theory:fidelity}
\mathcal{F}=\left(\frac{1}{2}+\Delta{X}_A^{2,out}\right)^{-\frac{1}{2}}\!\!\times\left(\frac{1}{2}+\Delta{P}_A^{2,out}\right)^{-\frac{1}{2}}.
\end{equation}
Assuming light and atoms to be initially shot noise limited we get a fidelity of $\mathcal{F}=\sqrt{{2}/{3}}\approx82\%$. 
The fidelity is limited by the initial noise of the $X_{A}^{in}$ spin component which was not canceled during the mapping process. If this quadrature would be squeezed before the pass of the light pulse the fidelity can approach unity.

It is known from earlier works \cite{Andersreview,Hammerer:2004je} that the coupling constant is related to the optical depth by $\kappa^{2} = d\,\eta$, where the optical depth is $d=N_{a}\sigma/A$, the scattering cross section is $\sigma = {3\lambda^{2}}/{2\pi}$ and the transversal area of the atomic sample is $A$. The coefficient $\eta\sim N_{p}/\Delta^{2}$ was called atomic depumping as it is closely related to a spin decay rate. The fact that the coupling constant depends on the decoherence rate is a very important observation. It means that for the great variety of quantum protocols based on such light--matter interaction a good performance is a compromise between the interaction strength and the decoherence. E.g. in order to reach a value of $\kappa=1$ in the memory protocol for a given optical depth $d$ a nonzero atomic depumping $\eta\sim 1/d$ is required.
By working with an ensemble with sufficiently large optical depth $d\gg 1$ the atomic depumping can be made small and this is the basis of the protocols for quantum interfaces based on the Faraday interaction which have been implemented in practice. In reality however, there is a limit to how large an optical depth that can be obtained in practice. Hence there will always be some atomic depumping for any implementation of a quantum information protocol. A proper optimization of such protocols requires knowledge of how exactly the depumping $\eta$ is related to the spin decay rates. In the next section we give the answer to this question. 

\section{Decoherence}\label{sec:decoherence}

\subsection{Example of decoherence}\label{subsec:decoherence-fidelity}

\begin{figure}[tbp]
\centering
\includegraphics[width=5.5cm]{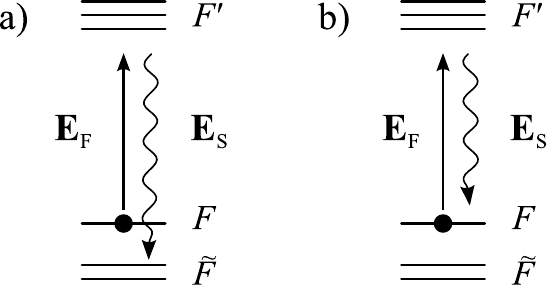}
\caption{Scheme of two possible decay processes.}
\label{fig:ways_of_spin_decay}
\end{figure}

Before presenting the rigorous treatment of the general case of spontaneous emission for multilevel atoms, let us include decay in a phenomenological basis and examine how an arbitrary decay would affect the atom light interaction. We expect that the spins and light would decay according to the Heisenberg-Langevin equations
	\begin{align}
	\label{light_decay}
	\frac{\partial}{\partial z}a_{\mu} &= -\frac{\gamma_{\mu}}{2L}a_{\mu} + F_{L,\mu}, &{\mu}&=\{x,y\},\\
	\label{spin_decay}
	\frac{\partial}{\partial t}j_i&=-\frac{\Gamma_i}T j_i + F_{i}, &i&=\{x,y,z\}.
	\end{align}
Since the decay process for light is due to scattering of the photons out of the mode of interest we shall see below that the added Langevin noise is simpler for light than the one for atoms. For light the noise is just the minimal noise required to preserve the commutation relation for the field. The noise is delta correlated in time and space and has a vanishing mean value
	\begin{align}
	{\langle}F_{L,\mu}^{\phantom{\dag}}(z,t)F_{L,\nu}^{\dag}(z',t'){\rangle} &=\frac{\gamma_{\mu}}L \delta_{\mu\nu}\delta(t-t')\delta(z-z'),\\
	{\langle}F_{L,\mu}(z,t){\rangle} &=0.
	\end{align}
The atomic spin decays in a number of ways. First, it can decay to another hyperfine level ($F-1$) and disappear from the interaction as it is show in Fig.~\ref{fig:ways_of_spin_decay}a. This is the same process which happens to photons. Another way for an atom to decay is to decay to the same hyperfine level as shown in Fig.~\ref{fig:ways_of_spin_decay}b. This results in essentially a random rotation of the original spin and it creates extra noise. We will derive the correlators of the spin noise operators in the following section, as well as the corresponding decay rates.

Using the definition of canonical variables for atoms \eqref{atomic_canonical_definition} we obtain the following expressions for the decay in canonical variables
	\begin{align}
	\frac{\partial}{\partial t}{X}_{A}&=\frac{1}{\sqrt{{\langle}j_x{\rangle}}}\frac{\partial}{\partial t}j_{y} - \frac{1}{2}\frac{j_y}{\sqrt{{\langle}j_x{\rangle}}}\cdot\frac{1}{{\langle}j_x{\rangle}} \frac{\partial}{\partial t}{\langle}{j}_x{\rangle}\nonumber\\
	\label{decay_canonical_atoms1}
	&=-\frac{\Gamma_{X}}T X_{A} + F_X(z,t),\\
	\label{decay_canonical_atoms2}
	\frac{\partial}{\partial t}{P}_{A}&=-\frac{\Gamma_{P}}T P_{A} + F_P(z,t).
	\end{align}
Here $\Gamma_{X(P)}=\Gamma_{y(z)}-\frac12\Gamma_{x}$, and $F_{X(P)}=F_{y(z)}/\sqrt{\langle j_{x}}\rangle$. Now we can consider equations of motion for the light-matter interaction in the presence of the decoherence in terms of the canonical variables
	\begin{align}
	\frac{\partial}{\partial z}\binom{X_{L}}{P_{L}}(z,t) &= \frac{\kappa(z,t)}{\sqrt{LT}}\binom{P_{A}}0\nonumber\\
	&\phantom{=}-\frac{\gamma_{y}}{2L} \binom{X_{L}}{P_{L}} + \binom{F_{L,X}}{F_{L,P}}(z,t),\\
	\label{th:atom_canon_evolution}
	\frac{\partial}{\partial t}\binom{X_{A}}{P_{A}}(z,t)&=\frac{\kappa(z,t)}{\sqrt{LT}}\binom{P_{L}}0\nonumber\\
	&\phantom{=}-\frac1T\binom{\Gamma_{X}X_{A}}{\Gamma_{P}P_{A}} + \binom{F_{X}}{F_{P}}(z,t).
	\end{align}
Here the coupling constant $\kappa(z,t)$ is a function of position and time due to the decay of the classical spin component and of the driving wave amplitude according to \eqref{light_decay},~\eqref{spin_decay}
	\begin{align}
	\kappa(z,t)&= \frac12 ga_{1}\sqrt{\langle j_{x}(z,t)\rangle} \langle a_{x}(z,t)\rangle\sqrt{LT}\nonumber\\
	&= \kappa\, e^{-\frac12(\gamma_{x}z/L+\Gamma_{x}t/T)}.
	\end{align}
Integration of the equations of motion over space and time gives us the input-output relations
	\begin{align}
	\label{theory:inout_big1}
	\binom{X_{L}^{out}}{P_{L}^{out}} &=
	\binom{X_{L}^{in}}{P_{L}^{in}} e^{-\frac{\gamma_{y}}2}
	+ \kappa_{L}\binom{P_{A}^{in}}{0} + \binom{F_{X_{L}}}{F_{P_{L}}},\\
	\label{theory:inout_big2}
	\binom{X_{A}^{out}}{P_{A}^{out}} &=
	\binom{X_{A}^{in}\,e^{-{\Gamma_{X}}}}{P_{A}^{in}\, e^{-{\Gamma_{P}}}} + \kappa_{A}\binom{P_{L}^{in}}{0} + \binom{F_{X_{A}}}{F_{P_{A}}}.
	\end{align}
Due to the space and time dependence of the function $\kappa(z,t)$ the integration of the equations of motion can be done by expanding light and atomic variables over Legendre polynomials in space and time domain \cite{Hammerer06,Vasilyev11}. 
However, we consider the decay rates to be small and therefore take into account only the zeroth order Legendre modes for atoms and light which are just integrals over length of the ensemble and over time of the interaction. The obtained relations are a generalization of \eqref{st:inoutL} and \eqref{st:inoutA} for an arbitrary decay. The following notations were used
	\begin{align}
	\kappa_{L} &= \kappa\, h\left(\frac{\Gamma_{x}+2\Gamma_{P}}2\right) h\left(\frac{\gamma_{x}-\gamma_{y}}2\right)e^{-\frac{\gamma_{y}}2},\\
	\label{theory:kappaa}
	\kappa_{A} &= \kappa\, h\left(\frac{\Gamma_{x}-2\Gamma_{X}}2\right) h\left(\frac{\gamma_{x}+\gamma_{y}}2 \right)e^{-\Gamma_{X}},\\
	&\qquad\qquad h(x)=\frac1x(1-e^{-x}).\nonumber
	\end{align}
The protocol of quantum memory based on the direct mapping requires $\kappa_{A}=1$, as described above. 
It is different from the constraint $\kappa=1$ without decoherence. 
For a given detuning an increase of the number of atoms leads to higher scattering for the signal field and an increase of the driving field strength gives rise to decay of atomic spins. Overall, it means that if the number of atoms for a given detuning is too low there is no way to meet the constraint by increasing the field power and vice versa.

The noise terms in the input output relations \eqref{theory:inout_big1} and \eqref{theory:inout_big2} are a bit lengthy to express exactly. 
One can show that the noise terms have the following approximate values
	\begin{align}
	\langle F_{X_{L}}^{2}\rangle &\approx \frac12 \gamma_{y} + \frac{\kappa^{2}}3\langle F_{P}^{2}\rangle,&
	\langle F_{P_{L}}^{2}\rangle &\approx \frac12 \gamma_{y},\\
	\langle F_{X_{A}}^{2}\rangle &\approx \langle F_{X}^{2}\rangle + \frac{\kappa^{2}}6\gamma_{y},&
	\langle F_{P_{A}}^{2}\rangle &\approx \langle F_{P}^{2}\rangle.
	\end{align}
Where the collective noise correlators for atoms are defined as follows
	\begin{equation*}
	\langle F_{X(P)}^{2}\rangle = \frac1{LT}\int\limits_{L}\!dzdz'\!\int\limits_{T}\!dtdt' \langle F_{X(P)}(z,t)F_{X(P)}(z',t')\rangle.
	\end{equation*}
In order to preserve mean values of the canonical variables during the mapping of the field state to the atoms one has to choose the feedback gain parameter to be $\nu = e^{\frac{\gamma_{y}}2}$. Assuming input states of light and atoms to be coherent we find the fidelity according to \eqref{theory:fidelity} which for low decay rates reads
	\begin{equation}
	\label{theory:decay-fidelity}
	\mathcal{F} \approx \sqrt{\frac23}\left(1 - \frac{11}{36}\gamma_{y} - \frac13\left[\langle F_{X}^{2}\rangle + 2\langle F_{P}^{2}\rangle-\Gamma_{X}\right]\right).
	\end{equation}
In order to evaluate the Fidelity we need to know all these decay rates and Langevin noises for a particular atomic ensemble. Here we come to the central part of the paper --- the proper treatment of spontaneous emission in multilevel atoms.


\subsection{General method}\label{sec:general_method}
In this section we find the full equations of motion from the single atom theory and assume that the formalism can be extended for a large collection of atoms that couple to their own reservoir and do not interact with each other. For a justification of this approach we refer to Ref. \cite{Martin}, where it is shown that the approximation is suitable for dilute elongated samples, for which $\rho\lambda^3\ll1$ and the Fresnel number of the ensemble is much larger than unity. The interaction Hamiltonian including both, forward and non-forward modes $\mathbf{E}=\mathbf{E}_{\rm{F}}+\mathbf{E}_{\rm{S}}$ is
	\begin{align}
	H_j&=\mathbf{E}^{(-)}{\boldsymbol\alpha}_j\mathbf{E}^{(+)}\nonumber\\
	&\simeq\mathbf{E}_{\rm{F}}^{(-)}{\boldsymbol\alpha}_j\mathbf{E}_{\rm{F}}^{(+)}+\mathbf{E}^{(-)}_{\rm{S}}{\boldsymbol\alpha}_j\mathbf{E}^{(+)}_{\rm{F}}+\mathbf{E}^{(-)}_{\rm{F}}{\boldsymbol\alpha}_j\mathbf{E}^{(+)}_{\rm{S}}\nonumber\\&=H^j_{\rm{coh}}+V_j,
	\end{align}
where $j$ labels the $j$-th atom. We have neglected the much weaker contribution $\mathbf{E}^{(-)}_{\rm{S}}{\boldsymbol\alpha}_j\mathbf{E}^{(+)}_{\rm{S}}$ which has no enhancement the by strong field in $\mathbf{E}_{\rm F}$. The coupling of forward modes to forward modes we identify as the
coherent interaction
	\begin{equation}
	H_{\rm{coh}} = \mathbf{E}^{(-)}_{\rm{F}}\boldsymbol\alpha\,\mathbf{E}^{(+)}_{\rm{F}}\equiv{}\frac{d_0^2}{\Delta}\mathbf{E}^{(-)}_{\rm{F}}\alpha\mathbf{E}^{(+)}_{\rm{F}}.
	\end{equation}
Here we have introduced dimensionless $\alpha$ which will be more convenient to use in what follows.
The interaction with the environment 
$\mathbf{E}_{\rm{S}}$ we will treat in Wigner-Weisskopf approximation. To do this we write the perturbation $V_j$ in the form
	\begin{equation}
	V_j=\frac{d_0^2}{\Delta}\sum_{\sigma}\int\frac{d^{3}k}{(2\pi)^{3}}\sqrt{\frac{\omega_k}{2\epsilon_0}}\boldsymbol{{\epsilon}}_{k\sigma}(b^{\dagger}_{k\sigma}{\alpha}_j\mathbf{E}_{\rm{F}}^{(+)}+\mathbf{E}_{\rm{F}}^{(-)}{\alpha}_jb_{k\sigma}).
	\end{equation}
The summation here is performed over all directions and polarizations of the non-forward electromagnetic modes. The resulting interaction will depend on the relative orientation of the atomic dipole moments hidden in the polarizability tensor $\alpha$ and the light field polarization vector. The net effect of this directional dependence appears as a difference in the decay rates and added noises for parallel and orthogonal light and atomic spin orientations. The modes in $\mathbf{E}_{\rm{S}}$ have a Hamiltonian
	\begin{equation}
	H_R=\sum_{\sigma}\int\frac{d^{3}k}{(2\pi)^{3}}\omega_kb^{\dagger}_{k\sigma}b_{k\sigma}.
	\end{equation}
We assume the system to start out in vacuum and proceed in a Wigner-Weisskopf approach by first finding the time evolution of the $b$-operators
	\begin{align}
	\frac{\partial}{\partial{t}}b_{k\sigma}(t) &= i[H_R+V_j,b_{k\sigma}]\nonumber\\
	&=-i\omega_kb_{k\sigma}(t)-i\frac{d_0^2}{\Delta}\sqrt{\frac{\omega_k}{2\epsilon_0}}\boldsymbol{\epsilon}_{k\sigma}{\alpha}_j(t)\mathbf{E}_{\rm{F}}^{(+)}(t),
	\end{align}
which has the formal solution
	\begin{align}
	 b_{k\sigma}(t) &= b_{k\sigma}(0)e^{-i\omega_kt}\nonumber\\
	&-i\frac{d_0^2}{\Delta}\sqrt{\frac{\omega_k}{2\epsilon_0}}\boldsymbol{\epsilon}_{k\sigma}\int_0^tdt'{\alpha}_j(t')\mathbf{E}_{\rm{F}}^{(+)}(t')e^{-i\omega_k(t-t')}.
	\end{align}
We are interested in the equation of motion of some operator $A$ (which can belong to either atoms or light). We insert the found expression for $\mathbf{E}_{\rm{S}}$ into $V_j$
\begin{widetext}
	\begin{multline}\label{big_equation}
	\frac{d}{dt}A(t) = 
	[H^j_{\rm{coh}},A] + 
	i\frac{d_0^2}{\Delta}\!\sum_{\sigma}\int\frac{d^{3}k}{(2\pi)^{3}}\sqrt{\frac{\omega_k}{2\epsilon_0}}\boldsymbol{\epsilon}_{k\sigma}
	\left(b^{\dagger}_{k\sigma}(0)e^{i\omega_k{t}}[{\alpha}_j\mathbf{E}_{\rm{F}}^{(+)},A](t)+[\mathbf{E}_{\rm{F}}^{(-)}{\alpha}_j,A](t)b_{k\sigma}(0)e^{-i\omega_kt}\right)\\
	-\Big(\frac{d_0}{\Delta}\Big)^2\sum_{\sigma}\!\int\!\frac{d^{3}k}{(2\pi)^{3}}\frac{\omega_k}{2\epsilon_0}\boldsymbol{\epsilon}_{k\sigma}^2\!\int_0^t\!dt'\!
	\left(\mathbf{E}_{\rm{F}}^{(-)}(t'){\alpha}_j(t')e^{i\omega_k(t-t')}[{\alpha_j}\mathbf{E}_{\rm{F}}^{(+)},A](t) 
	- [\mathbf{E}_{\rm{F}}^{(-)}{\alpha}_j,A](t){\alpha}_j(t')\mathbf{E}_{\rm{F}}^{(+)}(t')e^{-i\omega_k(t-t')}\right).
	\end{multline}
\end{widetext}
Going into the rotating frame of the forward field, which has the carrier frequency $\omega_0$ and performing the Markov approximation we may take the polarizability outside
the integral and find that the general equation of motion for an
observable $A$ is given by the quantum Langevin equation
	\begin{multline}
	\frac{d}{dt}A = i[H^j_{\rm{coh}},A] +\Big(\frac{d_0}{\Delta}\Big)^2
	\mathcal{L}_j(A)\\
	\label{decoh:heizenberg-langevin}
	+i\sqrt{\gamma}\frac{d_0}{\Delta}\left([\mathbf{E}_{\rm{F}}^{(-)}\alpha_j,A]\mathbf{f}_j + \mathbf{f}_j^{\dagger}[\alpha_j\mathbf{E}_{\rm{F}}^{(+)},A]\right).
	\end{multline}
Here $\gamma$ is the decay rate $\gamma = {d_0^2\omega_0^3}/{3\pi\epsilon_0}$ which is related to radiative decay rate of an atom with exited state electronic angular momentum $J'$ as $\gamma_{rad} = \gamma/(2J'+1)$. We have neglected the Lambshift, which is assumed to be incorporated in the transition frequency $\omega_0$, also we ignore collective effects in the limit of a dilute ensemble (see \cite{Martin} for details) as discussed in the beginning of Section~\ref{sec:general_method}. The decay is described by the Lindblad form
	\begin{multline}
	\label{decoh:heizenberg-langevin2}
	\mathcal{L}_j(A) = \frac{\gamma}{2}\left(2\mathbf{E}_{\rm{F}}^{(-)}{\alpha}_jA{\alpha}_j\mathbf{E}_{\rm{F}}^{(+)}\right.\\
	- \left.\mathbf{E}_{\rm{F}}^{(-)}{\alpha}_j^2\mathbf{E}_{\rm{F}}^{(+)}{A}-A\mathbf{E}_{\rm{F}}^{(-)}{\alpha}_j^2\mathbf{E}_{\rm{F}}^{(+)}	\right),
	\end{multline}
The decay gives two different contributions, one where the atom decays to the same state and will involve the $a$-coefficients belonging to $\alpha_{FF}$ and one to different states through $\alpha_{FF\pm1}$ and be described by $b$-coefficients. But we are not going to have any cross terms of the form $ab$ since they will oscillate at much higher frequencies corresponding to the hyperfine splitting $\sim$ GHz and average out to zero in \eqref{big_equation}.
Above we have introduced the noise operators,
	\begin{equation}
	\mathbf{f}_j(t) = \frac{d_0}{\sqrt{\gamma}}\sum_{\sigma}\int\frac{d^{3}k}{(2\pi)^{3}}\sqrt{\frac{\omega_k}{2\epsilon_0}}{\boldsymbol{\epsilon}_{k\sigma}}b^j_{k\sigma}(0)e^{-i(\omega_k-\omega_0)t}.
	\end{equation}
which in the Markov approximation are delta-correlated in time
	\begin{equation}
	\label{decoh:general_noise_operators}
	{\langle}[{f}_{i,\mu}(t),{f}_{j,\nu}^{\dagger}(t')]{\rangle} = \delta_{ij}\delta_{\mu\nu}\delta(t-t').
	\end{equation}
Below we shall apply these equations of motion to observables for both light and atoms. There is an essential difference between the decoherence of light and of atoms. The reason is that each time the atoms
undergo spontaneous emission they remain in the ensemble. For light
on the other hand a lost photon simply attenuates the beam, but there
is no change in the coherence of it. It means that we can model the
resulting noise for light as vacuum operators when considering the field $a,a^{\dagger}$.

\subsection{Light}
In the last section we showed how to treat the spontaneous emission and how to calculate the decay for the operators of interest. We will now apply this to the light field operators $a_{x}(z,t)$ and $a_{y}(z,t)$. First, one has to generalize the dynamics of a single atom studied above to a spatially extended ensemble.
%
The continuous equations of motion are obtained from the single atom version \eqref{decoh:heizenberg-langevin} by replacing operators with the space dependent ones defined in \eqref{theory:continuous_spin} and then integrating the right hand side over the ensemble volume.

\begin{figure}[tbp]
\centering
\includegraphics[width=8.5cm]{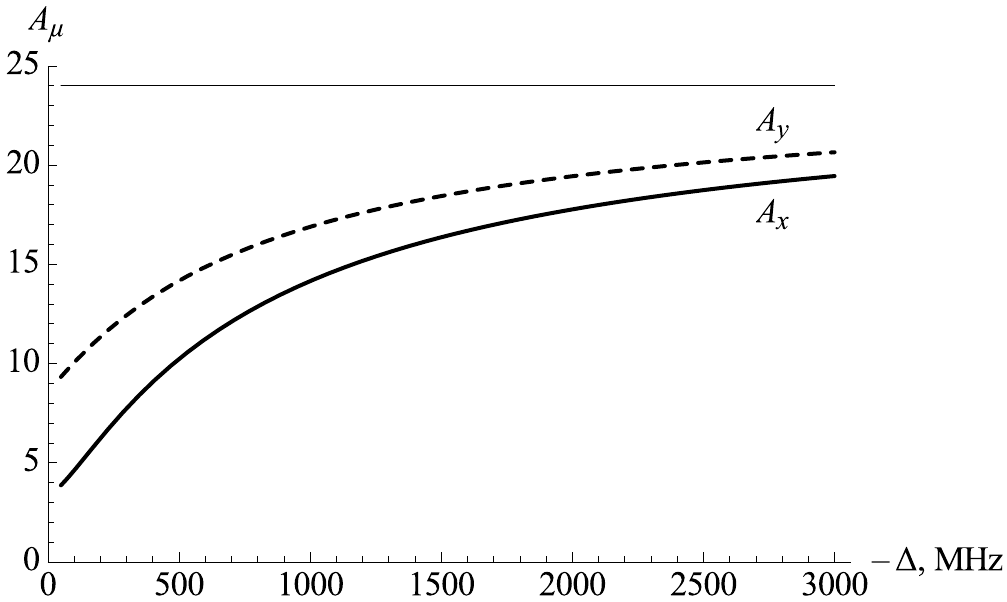}
\caption{Light attenuation and noise are described by the matrix $A_{\mu}$. Solid line shows the matrix element $A_{x}$ as a function of detuning $-\Delta$ in MHz and the dashed line corresponds to $A_{y}$. We assume atoms to be polarized along $x$. Cs in $F=4$.}
\label{fig:alpha_coeff}
\end{figure}

In order to obtain the required equations for the field propagation one has to consider the field evolution including the interaction Hamiltonian and the free field Hamiltonian. The resulting equation is obtained from the Heisenberg-Langevin equation~\eqref{decoh:heizenberg-langevin} by replacing $\frac{\partial}{\partial t} \rightarrow \frac{\partial}{\partial z} + \frac{\partial}{\partial t}$ as showed in Appendix~\ref{A:light-transform} and \cite{Kolobov99}. Since the light pulse is usually much longer than the size of the atomic ensemble we can omit the time derivative as it only describes retardation effect. Using the commutation relation for light field operators one obtains the following equations of motion
	\begin{align}
	\frac{\partial}{\partial z} a_{\mu}(z,t)\! &= i\int_{L} dz[H_{\rm{coh}},a_{\mu}(z,t)]\nonumber\\
	 &\phantom{=}- \frac{|g|\gamma}{4\Delta}\frac{N_{a}}L(\langle\alpha^{2}\rangle_{\mu\mu} a_{\mu}(z,t) \!+\! \langle\alpha^{2}\rangle_{\mu\nu} a_{\nu}(z,t))\nonumber\\
	&\phantom{=}+ F_{L,\mu}(z,t),\\
	 F_{L,\mu}(z,t) \!&=\! i\sqrt{\frac{|g|\gamma}{2\Delta}\frac{N_{a}}L}(\alpha_{\mu\mu}f_{\mu}(z,t) \!+\! \alpha_{\mu\nu}f_{\nu}(z,t)),
	\end{align}
where $\mu=\{x,y\}$ and $\mu\neq\nu$. Here we have assumed that the atoms are evenly distributed so that $\langle \alpha^{2}\rangle_{\mu\mu}(z) = \frac{N_{a}}L\langle\alpha^{2}\rangle_{\mu\mu}$ and the element $\langle\alpha^{2}\rangle_{ij}$ is to be understood as the $ij$'th element of the matrix $\alpha^{2}$. Using \eqref{decoh:general_noise_operators} we obtain the following properties of the averaged commutators of the noise operators $F_{L,\mu}$
	\begin{multline}
	\label{Langevin_light}
	\langle[F_{L,\mu}(z,t),F_{L,\nu}^{\dagger}(z',t')]\rangle = \\
	=\frac{|g|\gamma}{2\Delta}\frac{N_{a}}L\langle\alpha^{2}\rangle_{\mu\nu} \delta(t-t')\delta(z-z').
	\end{multline}
Since the matrix $\alpha^{2}$ depends on the atomic operators, the expressions derived here incorporates that the decoherence of the light field depends on the state of the atoms. To evaluate the formulas we therefore need to specify the atomic state. Here we assume that the atoms are initially polarized along the $x$-axis and evaluate the expressions in a coherent spin state. In principle there is also a possibility for cross-scattering between the $x$ and $y$ polarization, which is contained in the matrix element $\langle\alpha^2\rangle_{xy}$. In Appendix~\ref{app:alpha} we show, however, that this matrix element vanishes for this particular spin configuration. As noted above experiments sometimes use a different configuration with the atoms polarized perpendicular to the light polarization. Also in this case the cross polarization scattering vanishes, but this is not the case for a general orientation. Experimentally it is advantageous to avoid the cross polarization scattering since this avoids a rotation of the mean polarization of the light. Such rotation  may be hard to control precisely and this classical noise could therefore overwhelm the quantum noise around the mean which we are trying to control.  Therefore, the geometry with spin polarized parallel or orthogonal to the direction of light polarization is preferable. In both these cases we obtain the equation of motion for light in the form \eqref{light_decay} with the decay rates.
	\begin{align}\label{decay_of_light}
	\gamma_{\mu} &= \frac{|g|\gamma}{2\Delta}{N_{a}}\langle\alpha^{2}\rangle_{\mu\mu} = 
	2\frac{N_{a}}{N_{p}}\frac{\kappa^{2}}d A_{\mu},\\
	A_{\mu} &= \frac{\langle\alpha^{2}\rangle_{\mu\mu}}{Fa_{1}^{2}}.\nonumber
	\end{align}
Here the difference in the orientation of the atoms is again contained in the matrices $\langle\alpha^2\rangle_{\mu\mu}$. For Cesium with $F=4$ the $A_{\mu}$ coefficients are shown in Fig.~\ref{fig:alpha_coeff}. In the limit of large detuning the coefficients read
	\begin{equation}
	A_{x(y)} \to 24.
	\end{equation}
Taking into account coherent part of the interaction Hamiltonian we arrive at the full equations for the light field
	\begin{multline}
	\frac{\partial}{\partial z} a_{y}(z,t) = \frac{g}{2\sqrt T}\,a_{x}(z,t)\left[a_{1} j_{z}(z,t) + ia_{2}\{j_{x},j_{y}\}(z,t)\right]\\
	-\left[\frac{\gamma_{y}}2 - 2\,ia_{2}j_{y}^{2}(z,t)\right]a_{y}(z,t) + \sqrt{\gamma_{y}}f_{y}(z,t),
	\end{multline}
	\begin{multline}
	\frac{\partial}{\partial z} a_{x}(z,t) = \frac{-g}{2\sqrt T}\,a_{y}(z,t)\left[a_{1} j_{z}(z,t) - ia_{2}\{j_{x},j_{y}\}(z,t)\right]\\
	-\left[\frac{\gamma_{x}}2 - 2\,ia_{2}j_{x}^{2}(z,t)\right]a_{x}(z,t).
	\end{multline}
Since the $x$-polarized field is supposed to be strong and classical we neglect its quantum noise.

\subsection{Atoms}
In this section we consider atomic evolution in a similar way as it was done for light. To describe the decay of the single atomic spin we use expressions \eqref{decoh:heizenberg-langevin},~\eqref{decoh:heizenberg-langevin2} from above to obtain
\begin{align}
\frac{\partial}{\partial{t}}j_i(z,t)&=\Big(\frac{d_0}{\Delta}\Big)^2\mathcal{L}(j_i)(z,t),\\
\mathcal{L}(j_i)(z,t)
&=-\frac{\gamma}{2}\mathbf{E}^{(-)}_{\rm{F}}(z)[\alpha^2{j}_i+j_i\alpha^2-2\alpha{}j_i\alpha](z)\mathbf{E}^{(+)}_{\rm{F}}(z).
\end{align}
We expect this decay to be proportional to the total flux of photons and we can as a
good approximation ignore the position dependence of the light field,
allowing us to write for the collective spin
\begin{equation*}
\mathcal{L}(j_i)(t)
=-\frac{\gamma}{2}\mathbf{E}^{(-)}_{\rm{F}}\int{}dz[\alpha^2{j}_i+j_i\alpha^2-2\alpha{}j_i\alpha](z)\mathbf{E}^{(+)}_{\rm{F}}
\end{equation*}
The most notable contribution comes from the $x$ polarized part of
light. Since we already expect the effect of spontaneous emission to be small in the regime  where the quantum interface is operating (remember that it scales as ${1}/{\Delta^2}$) we can safely only consider the $xx$ element of the matrix
	\begin{equation*}
	\xi_i=\alpha^2j_i+j_i\alpha^2-2\alpha{j_i}\alpha.
	\end{equation*}
With this assumption we find that we can make the linearization
\begin{equation}
\mathcal{L}(j_i)(t)=-\frac{\gamma}{2}|E|^{2}\frac{N_p}{T}\Xi_i\,j_i(t),
\end{equation}
where the field amplitude is given by $|E|^{2}={\omega_{0}}/{2\epsilon_{0}A}$ and the decay rate coefficient $\Xi_i$ is defined through the
expression
\begin{equation}
\Xi_i\,j_i(t)=\int_0^L dz\,\xi_i(z,t).
\end{equation}
This element gives the magnitude of the decay of the respective
component of the spin. For the decay to a different hyperfine  level $F$ (the $b$ terms), we will neglect the term $2\alpha{j_i}\alpha$, which describe the increase in the
population of the final state. The reason is that once an atom decays to
another hyperfine level $F$ it will no longer be interesting for us, since we
restrict our analysis to collective behavior of many atoms in the
same $F$ state. This is reasonable since the
energy spacing between the two ground $F=I\pm\frac12$ hyperfine levels is big. For the same
reasons we obviously need to keep $2\alpha{j_i}\alpha$ for the
$a$-terms. Written in short form the decay reads
	\begin{equation}
	\frac{\partial}{\partial{t}}j_i(t)=-\frac{|g|\gamma}{4\Delta}\frac{N_p}{T}\Xi_i \,j_i(t) + {\rm Noise}.
	\end{equation}
This expression gives us the decay rates of the spin components defined in \eqref{spin_decay}. The noise terms will be treated in detail below. The decay rates depend on the relative orientation of spin and light polarizations as it is the case with the light decay. The expression for the decay rates now read
	\begin{align}
	\label{decay_of_atoms}
	\Gamma_{i_{\parallel(\!\perp\!)}} &= \frac{|g|\gamma}{4\Delta}{N_p}\,\Xi_{i_{\parallel(\!\perp\!)}} = 
	\frac{\kappa^{2}}d B_{i_{\parallel(\!\perp\!)}},\\
	B_{i_{\parallel(\!\perp\!)}} &= \frac{\Xi_{i_{\parallel(\!\perp\!)}}}{Fa_{1}^{2}}.\nonumber
	\end{align}
Similar to the discussion of spontaneous emission in section~\ref{subsec:Canonical-Operators} we have here expressed the atomic scattering in terms of the coupling constant $\kappa$ and the optical depth $d$. The details of the interaction and the difference between different atoms, which is our main interest here, is thus contained in the coefficient $B$. The required coefficients $\Xi$ are calculated in Appendix~\ref{app:xi} for an arbitrary atom. The $B_{i}$ coefficients for Cesium atoms are shown in Fig.~\ref{fig:Xi_coeff}, explicit expressions can be found in Appendix~\ref{app:explicit_expression}. In the limit of high detuning the coefficients for Cesium atoms with $F=4$ read
	\begin{align*}
	B_{x_{\parallel}} &\to {\frac{29}{2}},
	&B_{y_{\parallel}},\,B_{z_{\parallel}}&\to {\frac{25}{2}},\\
	B_{y_{\perp}}&\to {\frac{37}{4}},
	&B_{x_{\perp}},\,B_{z_{\perp}}  &\to \frac12B_{x_{\parallel}} .
	\end{align*}
The remaining term in the Heisenberg-Langevin equation \eqref{decoh:heizenberg-langevin} gives us the spin noise components
\begin{figure}[tbp]
\centering
\includegraphics[width=8.5cm]{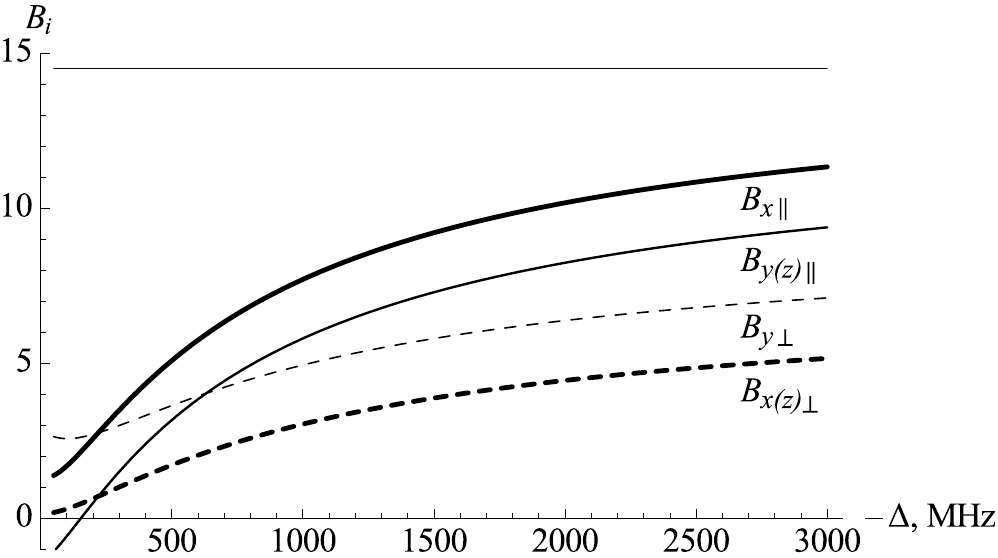}
\caption{Spin decay coefficients $B_{i}$ as functions of detuning $-\Delta$ in MHz for parallel configuration (spin and light are $x$ polarized) are shown by thick solid line for $x$ spin component and by thin solid curve for $y$ and $z$ spin projections. The orthogonal configuration (spin is $x$ polarized and the light is $y$ polarized) is presented by the dashed curves, thick line is $x$ and $z$, thin is $y$ spin component. The plot is calculated for Cesium atoms with ground state $F=4$.}
\label{fig:Xi_coeff}
\end{figure}
	\begin{equation*}
	\frac{\partial}{\partial{t}}j_i(z,t)=i\sqrt{\gamma}\frac{d_0}{\Delta}\left(\mathbf{E}^{(-)}_F	 [\alpha,j_i]\mathbf{f}+\mathbf{f}^{\dagger}[\alpha,j_i]\mathbf{E}^{(+)}_F\right).
	\end{equation*}
From here the noise operators for canonical spin density variables defined above as $F_X(z,t)$ and $F_P(z,t)$ are given by
	\begin{equation}
	F_{X(\!P)}(z,t)\! =\! {\frac{i\sqrt\gamma d_{0}}{\Delta\sqrt{n\langle j_x^{a} \rangle}}}\!\left(\!{\mathbf{E}}^{(-)}_F[\alpha_,j_{y(z)}]\mathbf{f} \!+\! \mathbf{f}^{\dagger}[\alpha,j_{y(z)}]{\mathbf{E}^{(+)}_F}\!\right)\!.
	\end{equation}
These are the noise operators introduced in \eqref{th:atom_canon_evolution}. The collective noise operators are obtained by averaging it over the light pulse duration and the length of the atomic ensemble
	\begin{equation}
	\label{langevin_spinX_definition}
	F_{X(P)} = \frac1{\sqrt{LT}}\int_{L}dz\int_{T}dt\,F_{X(P)}(z,t).
	\end{equation}
When forming the combinations such as ${\langle}F_XF_X{\rangle}$ then, since we have assumed that there are no photons in the reservoir with frequency at $\omega_0$, the only combination which survives is ${\langle}\mathbf{f}\,\mathbf{f}^{\dagger}{\rangle}$. Moreover, we have made the assumption that the $x$ (or $y$ in case of orthogonal configuration of light and spin polarizations) component of the light field is dominating, so we will only have to consider the $xx$($yy$) element of the matrix $\zeta_{i}^{2}=(i[\alpha,j_i])^2$. 
So ultimately what we will have left is
	\begin{align}
	\label{langevin_spinX}
	{\langle}F_{X(P)}^{2}{\rangle}_{\parallel(\!\perp\!)} &=\frac{|g|\gamma}{2\Delta}		\frac{N_p}{F}\langle \zeta^{2}_{y(z)}\rangle_{\parallel(\!\perp\!)} = 
	\frac2F \frac{\kappa^{2}}d C_{y(z)_{\parallel(\!\perp\!)}},\\
	C_{i_{\parallel(\!\perp\!)}} &= \frac{\langle \zeta^{2}_{i}\rangle_{\parallel(\!\perp\!)}}{(Fa_{1})^{2}}.\nonumber
	\end{align}
The different $\langle\zeta^{2}\rangle$ are calculated in Appendix~\ref{app:atomic_noise} and are generally quite complicated. Since any orientation besides the parallel and orthogonal will give too much noise on the light as discussed above, we only consider these two settings. The spin noise coefficients $C_{i}$ required for the calculation of canonical spin variables noises are shown in Fig.~\ref{fig:g2_coeff}, the explicit expressions are given in Appendix~\ref{app:explicit_expression}. In the limit of high detuning the coefficients for Cesium atoms with $F=4$ read
	\begin{align*}
	C_{y_{\parallel}},\,C_{z_{\parallel}} \to \frac{29}{2},\qquad	C_{y_{\perp}} \to {\frac{53}{4}},\qquad
	C_{z_{\perp}} \to {\frac{37}{4}}.
	\end{align*}
The cross correlations of the Langevin noise are closely related to the spin decay rates as shown in Appendix~\ref{app:spin_deacy_to_noise_relation}. The commutator for noise reads
	\begin{equation}
	\label{theory:spin_deacy_to_noise_relation}
	{\langle}[F_{X},F_{P}]{\rangle}_{\parallel(\!\perp\!)} =
	\frac{i}F (\Gamma_{y} + \Gamma_{z} - \Gamma_{x})_{\parallel(\!\perp\!)}.
	\end{equation}
Moreover, one can show that the anticommutator of the noise terms is equal to zero.
\begin{figure}[tbp]
\centering
\includegraphics[width=8.5cm]{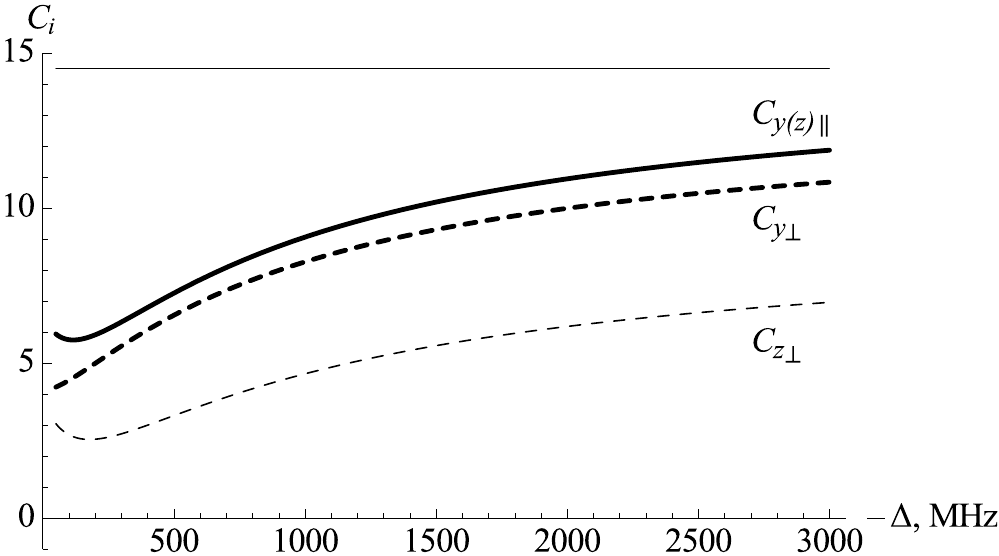}
\caption{Spin noise coefficients $C_{i}$ as functions of detuning $-\Delta$ in MHz for Cesium with ground state $F=4$. The solid line represents $y$ and $z$ noise coefficients for $x$ polarized atoms and light. The dashed curves correspond to the orthogonal configuration of atomic spin and light. The thick dashed curve shows $C_{y_{\perp}}$ and the thin dashed line is $C_{z_{\perp}}$.}
\label{fig:g2_coeff}
\end{figure}


\section{Summary}\label{sec:summary}

In this section we summarize the obtained results and as an example we apply the results to determine the decay rates and Langevin noises for the specific case of Cesium atoms. First of all, the decay of the spins and light is described by the Heisenberg-Langevin equations
	\begin{align*}
	\frac{\partial}{\partial z}a_{\mu} &= -\frac{\gamma_{\mu}}{2L}a_{\mu} + F_{L,\mu}, &{\mu}&=\{x,y\},\\
	\frac{\partial}{\partial t}j_i &=-\frac{\Gamma_i}T j_i + F_{i}, &i&=\{x,y,z\}.
	\end{align*}
{As it was mentioned earlier there are two preferable configurations available. Either the light polarization is oriented along the same axis as the spins of the atoms ($x$) or the light is orthogonal to the atomic polarization ($y$).}
In these cases the decay rates for light are given by \eqref{decay_of_light} and the atomic decay rates can be found in \eqref{decay_of_atoms}. They read
	\begin{align*}
	\gamma_{\mu} &= 	2\frac{N_{a}}{N_{p}}\frac{\kappa^{2}}d A_{\mu},
	&\Gamma_{i_{\parallel(\!\perp\!)}} &= \frac{\kappa^{2}}d B_{i_{\parallel(\!\perp\!)}},\\
	A_{\mu} &= \frac{\langle\alpha^{2}\rangle_{\mu\mu}}{Fa_{1}^{2}}, 
	&B_{i_{\parallel(\!\perp\!)}} &= \frac{\Xi_{i_{\parallel(\!\perp\!)}}}{Fa_{1}^{2}}.
	\end{align*}
From these expression we see that one can obtain a strong interaction $\kappa\sim1$ with negligible decoherence provided a that the atomic ensemble has a sufficiently high optical depth $d\gg 1$. Furthermore the decoherence of the light modes can be suppressed relative to the atomic decay provided a larger number of photons than atoms is used $N_P\gg N_A$. Most of these expressions are well known from previous work in the field \cite{Andersreview}. The main new result in this work are the exact expressions for the coefficients $A$, $B$ and $C$, which are numbers of order unity, which contain the information specific to the particular atom one is considering. 
We will evaluate the required coefficients for $^{133}$Cs optically pumped to ground state $F=4$. In this case we have $I=7/2$, $J=1/2$, $J'=3/2$, $F=4$, $\tilde F=3$. The coefficient $a_{1}$ is given by \eqref{a_coefficients_all}, which is evaluated for $F=\tilde F=4$ and $k=1$
	\begin{equation}
	a_{1} = \frac{7}{5760}\left(\frac{176}7 - \frac{3}{1-\frac{\Delta_{45}}{\Delta}} - \frac{5}{1-\frac{\Delta_{35}}{\Delta}}\right).
	\end{equation}
The matrix $\langle\alpha^{2}\rangle$ whose diagonal components enter in $A_{\mu}$ can be  found in Appendix~\ref{app:alpha}. The Cartesian components of the matrix for the spin oriented along $x$-axis are expressed via its spherical components by \eqref{app:alpha_dekartX} and \eqref{app:alpha_dekartY}. The required spherical components of the matrix are given by \eqref{app:alpha_spherical}. One needs to know the coefficients $a_{k}^{F\tilde F}$ mentioned above and the coefficients $c_{k}$ given by \eqref{app:coeff_ck}. The result for Cs atoms is
	\begin{align*}
	\langle\alpha^{2}\rangle_{xx} &= \frac1{80}\left(1 + \frac7{3(1-\frac{\Delta_{45}}	{\Delta})^{2}}\right) \to \frac1{24},\\
	\langle\alpha^{2}\rangle_{yy} &= \frac1{720}\left(\! 23 + \frac{21}{8(1-\frac{\Delta_{45}}{\Delta})^{2}} + \frac{35}{8(1-\frac{\Delta_{35}}{\Delta})^{2}}\!\right) \to \frac1{24}.
	\end{align*}
Since the spin is considered to be oriented along $x$, the $xx$-element of the scattering matrix represents decay of the field polarized along the spin, and the $yy$-component is correspondingly the decay rate coefficient for the orthogonally polarized field configuration. 

Calculation of the spin decay coefficients $\Xi$ entering the $B$ coefficients is similar but a bit more lengthy since it involves three spin projections and two possible configurations. The basic expressions for parallel and orthogonal configurations of spin and light are given by \eqref{app:decay_spin_dekart1}--\eqref{app:decay_spin_dekart2}. The required spherical components of $\langle\xi_{\mu} j_{\nu}\rangle$ are defined in \eqref{app:decay_spin_spherical1}--\eqref{app:decay_spin_spherical3}. The definitions involve the same coefficients $a_{k}^{F\tilde F}$ and $c_{k}$ used above for the evaluation of the light decay rates. For reference we provide the result of the calculation for the expression $\langle\Xi_{y}\rangle_{\parallel}$ for Cesium atoms $F=4$.
	\begin{multline*}
	\langle\Xi_{y}\rangle_{\parallel} = \frac{7}{1152}\left(-\frac{152}{175}
	-\frac{7}{12 \left(1-\frac{\Delta_{35}}{\Delta}\right)
   \left(1-\frac{\Delta_{45}}{\Delta}\right)}\right.\nonumber\\
   + \left. \frac{1}{2\left(1-\frac{\Delta_{35}}{\Delta}\right)^2}
   +\frac{76}{75\left(1-\frac{\Delta_{45}}{\Delta}\right)}
   +\frac{351}{100\left(1-\frac{\Delta_{45}}{\Delta}\right)^2}\right).
	\end{multline*}

The Langevin noise operators for the light modes \eqref{Langevin_light} are simply the minimal noise required to preserve the commutation relation for the free field.
	\begin{equation*}
	{\langle}F_{L,\mu}^{\phantom{\dag}}(z,t)F_{L,\nu}^{\dag}(z',t'){\rangle} =	\frac{\gamma_{\mu}}L \delta_{\mu\nu}\delta(t-t')\delta(z-z').
	\end{equation*}

In the context of quantum information protocols it is more convenient to study the evolution of the canonical variables for the spin. In this case the decay process is described according to \eqref{decay_canonical_atoms1} by 
	\begin{align*}
	\frac{\partial}{\partial t}{X}_{A}&=-\frac{\Gamma_{X}}T X_{A} + F_X(z,t),\\
	\frac{\partial}{\partial t}{P}_{A}&=-\frac{\Gamma_{P}}T P_{A} + F_P(z,t).
	\end{align*}
Here $\Gamma_{X(P)}=\Gamma_{y(z)}-\frac12\Gamma_{x}$, and $F_{X(P)}=F_{y(z)}/\sqrt{\langle j_{x}}\rangle$. 
The Langevin noise correlator for the collective canonical variables is found in \eqref{langevin_spinX_definition} and \eqref{langevin_spinX} and it reads
	\begin{equation*}
	{\langle}F_{X(P)}^{2}{\rangle}_{\parallel(\!\perp\!)} =
	\frac2F\frac{\kappa^{2}}d {C_{y(z)_{\parallel(\!\perp\!)}}},\qquad
	C_{i_{\parallel(\!\perp\!)}} = \frac{\langle \zeta^{2}_{i}\rangle_{\parallel(\!\perp\!)}}{Fa_{1}^{2}},
	\end{equation*}
The averaged commutator for the noise is given by \eqref{theory:spin_deacy_to_noise_relation}:
	\begin{equation*}
	{\langle}[F_{X},F_{P}]{\rangle}_{\parallel(\!\perp\!)} =
	\frac{i}F(\Gamma_{X}+\Gamma_{P})_{\parallel(\!\perp\!)}.
	\end{equation*}
To evaluate this expression one needs to calculate the matrix $\langle \zeta_{i}^{2}\rangle$. The calculation of this is essentially the same as for the spin decay coefficient. The expressions for the Cartesian components via the spherical one are given by \eqref{app:langevin_dekart1}--\eqref{app:langevin_dekart2}. Spherical components for different $F$ and $\tilde F$ are found in \eqref{app:langevin_spherical1}--\eqref{app:langevin_spherical3}. For reference we provide the result of the calculation for the expression $\langle \zeta_{y}^{2}\rangle_{\parallel}$ for Cesium atoms with $F=4$.
	\begin{multline*}
	\langle \zeta_{y}^{2}\rangle_{\parallel} = \frac7{2400}\left(\frac{176}{21}
	-\frac{175}{72 \left(1-\frac{\Delta_{35}}{\Delta }\right)
   \left(1-\frac{\Delta_{45}}{\Delta }\right)}\right.\\
   +\left. \frac{25}{12 \left(1-\frac{\Delta_{35}}{\Delta }\right)^2}
   -\frac{88}{9 \left(1-\frac{\Delta_{45}}{\Delta }\right)}
   +\frac{83}{8\left(1-\frac{\Delta_{45}}{\Delta }\right)^2}\right).
	\end{multline*}
Expressions for all of the coefficients $A,\,B,\,C$ for $^{133}$Cs are explicitly given in Appendix~\ref{app:explicit_expression}.

\subsection{Example of application}

\begin{figure}[tbp]
\centering
\includegraphics[width=8.5cm]{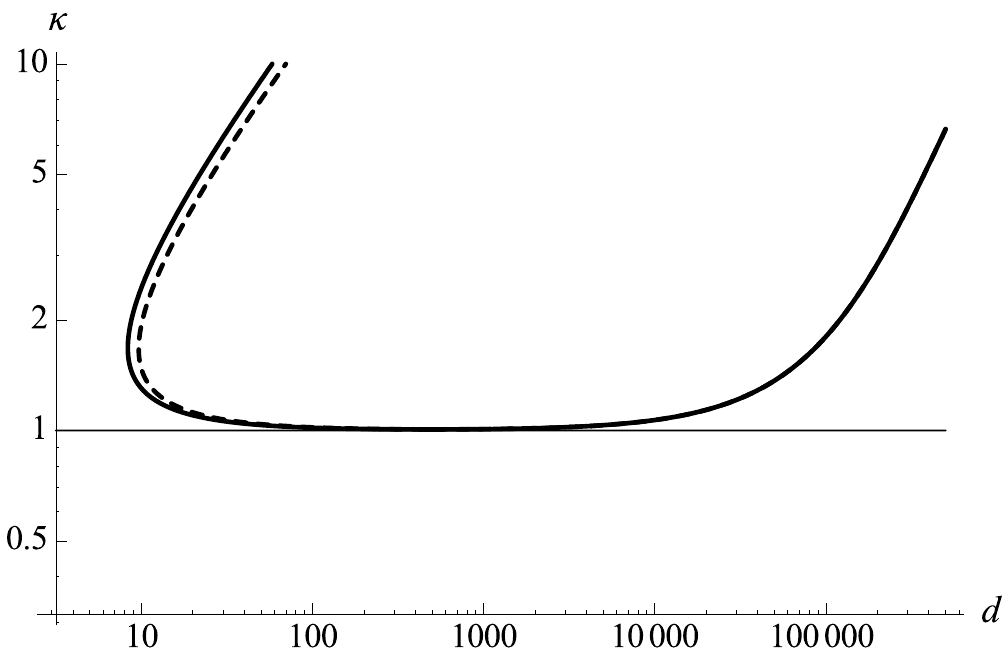}
\caption{Solution of the equation $\kappa_{A}=1$ as a function of the optical depth $d=N_{a}\sigma/A$. The thin solid line corresponds to the constraint without decoherence $\kappa=1$. Thick solid and dashed curves represent respectively the solutions for the parallel and orthogonal atomic spin and light polarization configurations. The plot is calculated for Cesium atoms with ground state $F=4$ and detuning $\Delta=-500\,$MHz.}
\label{fig:kappa_eq_one}
\end{figure}

Now we can use the derived expressions for the light and and atomic decay rates to evaluate  the Fidelity obtained in section~\ref{subsec:decoherence-fidelity}. First, let us consider the coupling constant $\kappa_{A}$ given by \eqref{theory:kappaa} for the case of $^{133}$Cs. In Fig.~\ref{fig:kappa_eq_one} we show the solution of the equation $\kappa_{A}=1$ which gives us the required parameter regime for the quantum memory based on direct mapping protocol. One can see that within a wide range of values for the optical depth $d=N_{a}\sigma/A$ the requirement $\kappa_{A}=1$ coincides with the condition $\kappa=1$. However, when it reaches  extreme values the decay processes significantly affect the system evolution and one is required to provide $\kappa>1$ in order to fulfill the quantum memory condition for transfer of the the mean amplitude.

For a given optical depth $d$ one can express the fidelity \eqref{theory:decay-fidelity} by the approximation of small decay probability. Consequently for the parameter regime where $\kappa_{A}=1$ coincides with the constriction $\kappa=1$ we have
	\begin{align}
	\mathcal{F} &\approx \sqrt{\frac23}\left(1 - c_{L}\left(\frac{\gamma}{\Delta}\right)^{2}d - c_{A}\,\frac{1}d\right)\\
	c_{L} &= \frac{11}{12}\frac1{Fa_{1}^{2}}{\langle\alpha^{2}\rangle}\\
	c_{A} &= \frac2{3(Fa_{1})^{2}}\left[\langle \zeta_{y}^{2}\rangle + 2\langle \zeta_{z}^{2}\rangle - \frac{F}2(\Xi_{y}\! -\frac12\Xi_{x})\right]
	\end{align}
In the limit of $\Delta\to-\infty$ we can neglect the noise on the light the coefficients and the fidelity read
	\begin{align}
	\mathcal{F} &\to \sqrt{\frac23}\left(1 - c_{A}\,\frac{1}d\right),\\
	c_{A_{\parallel(\perp)}} &\to \frac{11}2\left(\frac{41}{12}\right).
	\end{align}
{In agreement with Ref. \cite{Andersreview} the deviation of the fidelity from the ideal one is inversely proportional to the optical depth for large detunings. The results of the paper allow us to identify the coefficient of this proportionality. It shows an advantage of the orthogonal configuration of light and spin polarizations over the parallel one in the particular quantum memory protocol discussed here for illustration.}
	
\begin{figure}[tbp]
\centering
\includegraphics[width=8.5cm]{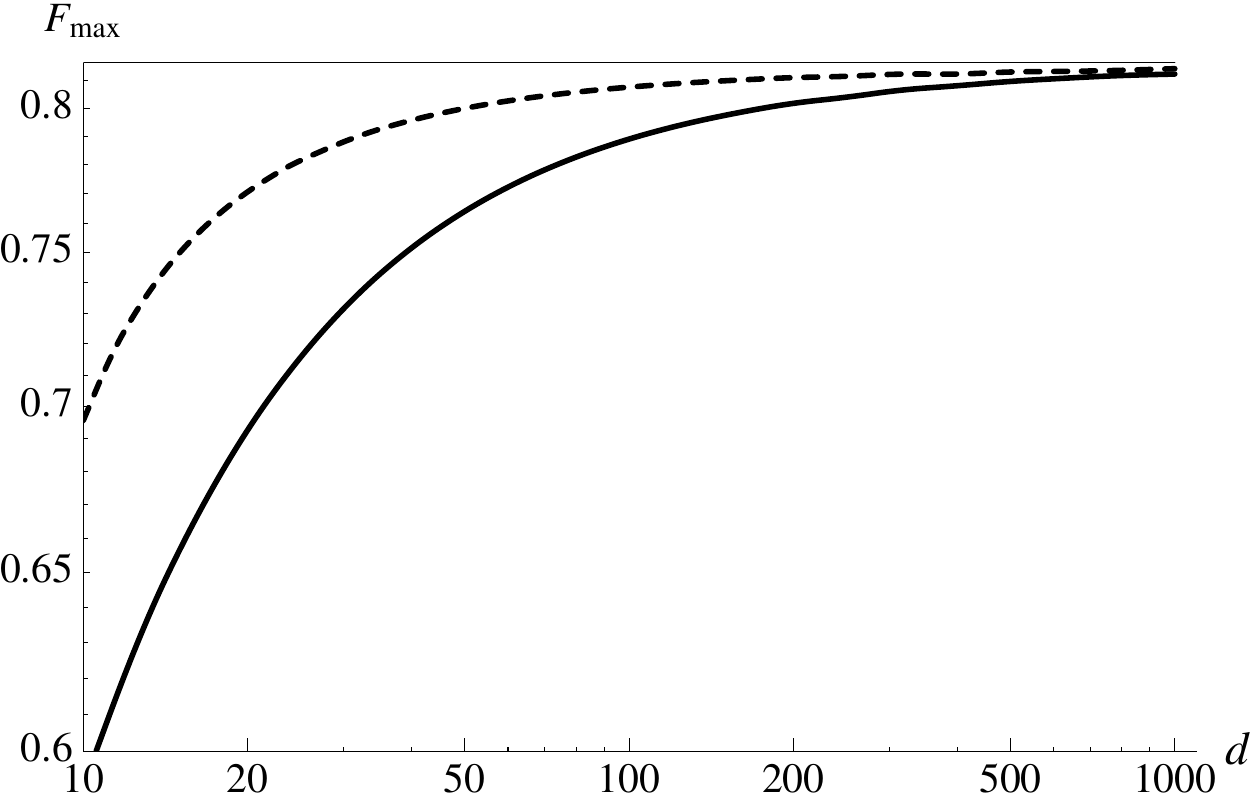}
\caption{Optimal fidelity as a function of optical depth for Cesium with $F=4$. Solid and dashed lines correspond respectively to the  parallel and orthogonal configurations of atomic spin and light polarization. Thin solid line on top shows the limiting value $\sqrt{2/3}$.}
\label{fig:optimal_fidelity_dA}
\end{figure}

\begin{figure}[tbp]
\centering
\includegraphics[width=8.5cm]{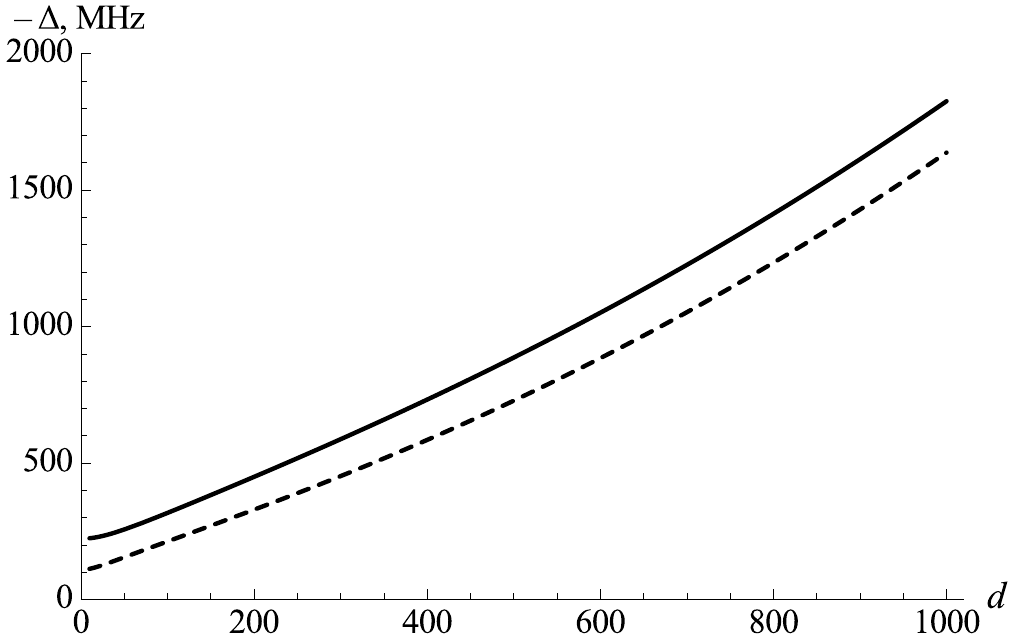}
\caption{Optimal detuning as a function of optical depth for Cesium with $F=4$. Solid and dashed lines correspond respectively to the  parallel and orthogonal configurations of atomic spin and light polarization.}
\label{fig:optimal_detuning}
\end{figure}

\begin{figure}[tbp]
\centering
\includegraphics[width=8.5cm]{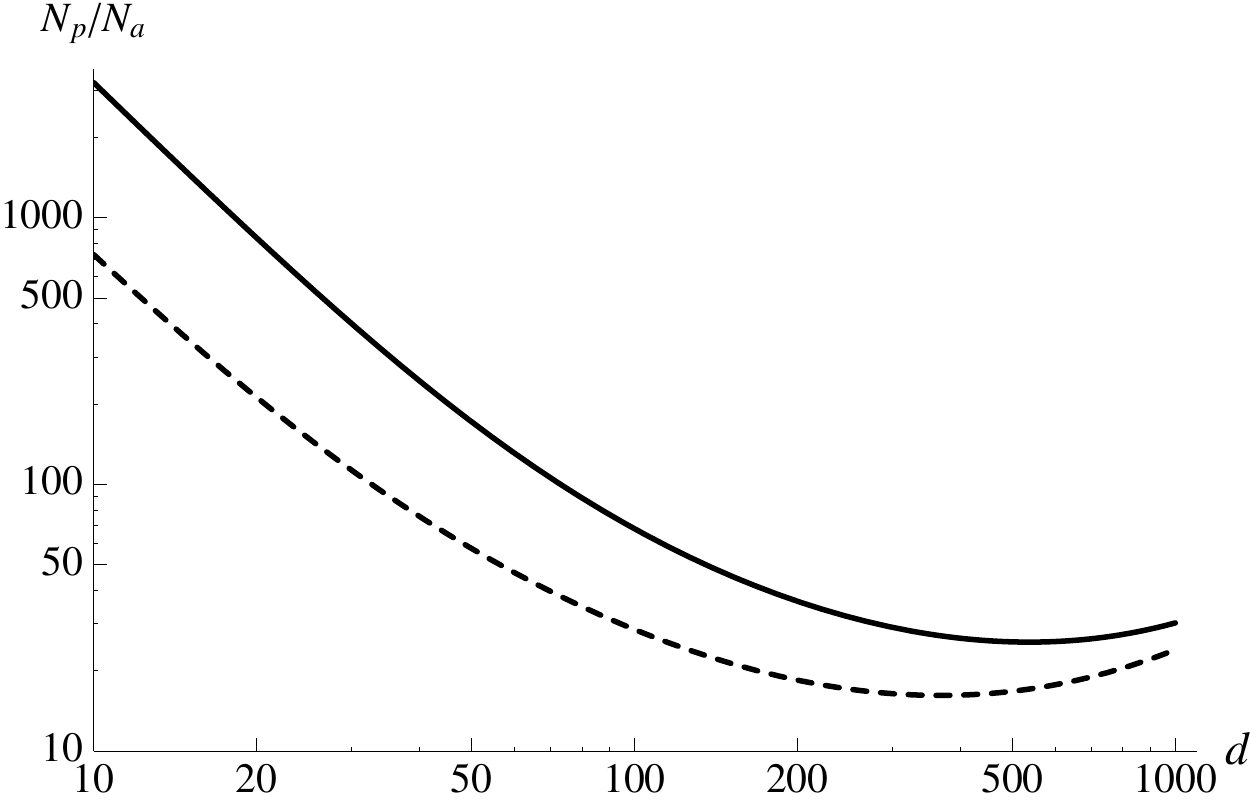}
\caption{Optimal ratio of photons to atoms as a function of optical depth for Cesium with $F=4$. Solid and dashed lines correspond respectively to the  parallel and orthogonal configurations of atomic spin and light polarization.}
\label{fig:optimal_ratio}
\end{figure}

\begin{figure}[tbp]
\centering
\includegraphics[width=8.5cm]{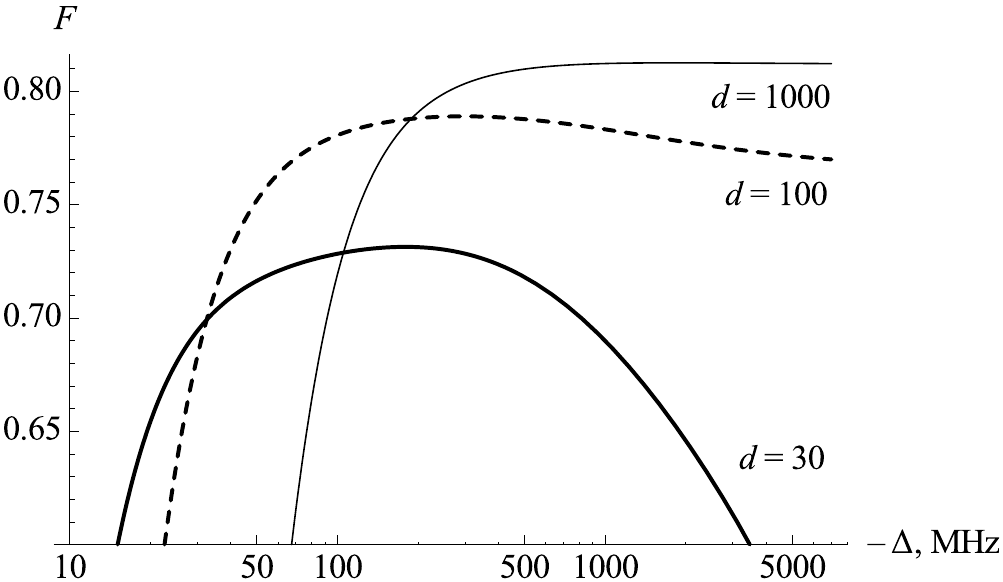}
\caption{Fidelity for different optical depths as function of detuning for Cesium with $F=4$. Solid, dashed, and thin curves correspond to optical depths $d=30,\,100,\,1000$ respectively. The calculation is done for the  parallel configuration of atomic spin and light polarization.}
\label{fig:fidelities_of_detuning}
\end{figure}

For a finite detuning the fidelity can be optimized by varying the ratio of photon number to the number of atoms and by changing the detuning from the resonance. The resulting optimal fidelity as a function of optical depth is shown in Fig.~\ref{fig:optimal_fidelity_dA}.
The corresponding optimal detuning and optimal ratio of photons to atoms as functions of the optical depth are shown in Fig.~\ref{fig:optimal_detuning} and Fig.~\ref{fig:optimal_ratio} respectively. One has to take into account that these 
are very flat maxima, and hence the precise value of the detuning is not that important. This is illustrated in Fig.~\ref{fig:fidelities_of_detuning} where fidelities for several optical depths are shown as a function of detuning. Furthermore one should also bear in mind that experimentally other considerations  such as Doppler broadening may be important. Furthermore to illustrate our method we have for simplicity neglected the tensor part of the coherent interaction, and this may alter the conclusion reached here.


\section{Conclusion}
In this paper we have found in accordance with earlier work that spontaneous emission occurs faster for parallel polarizations than perpendicular --- a consequence of selection rules and Clebsch--Gordan coefficients. We have also found the precise decay rates and noise correlations that take into account the full level structure of the atom and our procedure can easily be applied to similar systems. 
In this paper we have given a comprehensive discussion of quantum noise from spontaneous emission for  light matter quantum interfaces  based on the Faraday effect. Taking into account the full level structure of the atoms we derive for the first time the full expressions for the decay and quantum noise arising from spontaneous emission.  In agreement with previous treatments based on simplified models our theory shows that the decay and noise can be made to vanish for a sufficiently large optical depth of the ensemble. Given that experiments will always work with a finite optical depth it is, however, important to have a detailed understanding of the noise. In particular,  in view of the rapid experimental advances a thorough understanding of fundamental noise sources will be crucial for further increasing the efficiency of quantum memories and for assessing the feasibility of new advances for light matter quantum interfaces. The tradeoff between the coherent dynamics and unavoidable fundamental losses, investigated in full detail here, will set the ultimate limits to the  performance of future quantum networks. 

\begin{acknowledgments}
We acknowledge support from the EU project MALICIA, funding through the Centre for Quantum Engineering and Space-Time Research (QUEST) at the Leibniz University Hanover, and the Lundbeck foundation.
\end{acknowledgments}


\appendix

\section{Adiabatic elimination}\label{app:adiabat-elim}
{\noindent}The projections defined in the section \ref{sec:interaction} will be used for our treatment of the
operators and if we specifically apply it to the dipole-operator $\mathbf{d}$
we get (Note that we do not get any contributions from the terms $P_g\mathbf{d}P_g,P_e\mathbf{d}P_e$ since
$\mathbf{d}$ is a parity odd operator.)
\begin{align}
\mathbf{d}&=(P_g+P_e)\mathbf{d}(P_g+P_e)=P_e\mathbf{d}P_g+P_g\mathbf{d}P_e\nonumber\\&={\mathbf{d}}^{(+)}+{\mathbf{d}}^{(-)}.
\end{align}
Here $\mathbf{d}^{(+)}$ is the raising operator raising the atoms from the ground state $|F,m{\rangle}$ to the excited state $|F',m'{\rangle}$.
In the following we will go into the rotating frame with respect to
the laser frequency $\omega_0$, so the energy of the atom is described by the Hamiltonian
\begin{equation}
H_{{A}}=\sum_{F'}\Delta_{F'}P_{F'},
\end{equation}
where $\Delta_{F'}$ is the detuning of the exited stated from the
laser frequency.
To ease notation we treat by both using a symbol $g$ or $e$ to represent respectively ground states $|F,m{\rangle}$ and exited states $|F',m'{\rangle}$.
\begin{equation}
H_{\rm{int}}=-\sum_{ge}\mathbf{E}^{(-)}\mathbf{d}^{(-)}_{ge}|g{\rangle}{\langle}e|+\rm{h.c.}
\end{equation}
Taking the commutator with $H_{\rm{A}}+H_{\rm{int}}$ we get the following equation
	\begin{multline}
	\frac{d}{dt}|g{\rangle}{\langle}e|=-i\Delta_e|g{\rangle}{\langle}e|\\
	+ i\mathbf{E}^{(+)}(\sum_{g'}\mathbf{d}^{(+)}_{eg'}|g{\rangle}{\langle}g'|-\sum_{e'}\mathbf{d}^{(-)}_{ge'}|e'{\rangle}{\langle}e|).
	\end{multline}
So in the adiabatic regime we neglect the time derivative and obtain
\begin{equation}
|g{\rangle}{\langle}e|=\sum_{g'}\frac{\mathbf{E}^{(+)}\cdot\mathbf{d}^{(+)}_{eg'}}{\Delta_e}|g{\rangle}{\langle}g'|.
\end{equation}
Inserting this into the Hamiltonian we get the expression
\begin{align}
	 H_{\rm{int}}^{\rm{eff}}&=H_{\rm{int}}+H_{\rm{A}}=\mathbf{E}^{(-)}\cdot{\boldsymbol\alpha}\cdot\mathbf{E}^{(+)}\nonumber\\
	& {\,\,\equiv}\,H_{\rm{int}},\\
	{\boldsymbol\alpha} &= -\sum_{g'ge}\frac{\mathbf{d}^{(-)}_{g'e}\mathbf{d}^{(+)}_{eg}}{\Delta_e}|g'{\rangle}{\langle}g|
\end{align}
Here $\mathbf{d}^{(+)}_{g'e}\mathbf{d}^{(-)}_{eg}\sim\mathbf{d}^{(+)}_{g'e}\wedge\mathbf{d}^{(-)}_{eg}$ is the dyadic vector product of the dipole operator with itself.

\section{Construction of the Hamiltonian}
\label{A:hamiltonian_constraction}

The dipole moment $\mathbf{d}$ can be expanded in spherical components by
	\begin{equation}
	\mathbf{d} = \sum_{q}d_{q}^{\vphantom{*}}\mathbf{e}_{q}^{{*}} = \sum_{q}(-1)^{q}d_{-q}\mathbf{e}_{q}.
	\end{equation}
The component of $\mathbf{d}$ for the $F\to F'$ transition can thus be expressed as
	\begin{equation}
	\mathbf{d}^{(+)}_{F'F} = \sum_{q,m,m'}\langle F'm' |d^{(+)}_{q}|Fm\rangle|F'm'\rangle\langle Fm|\mathbf{e}_{q}^{*}.
	\end{equation}
The Wigner-Eckart theorem states that the matrix elements can be expressed as 
	\begin{multline}
	\langle F'm' |d^{(+)}_{q}|Fm\rangle = C_{Fm\,1q}^{F'm'}\langle F'\|d\|F\rangle\\
	= (-1)^{F-1+m'}\sqrt{2F'+1}\tj{F}{1}{F'}{m}{q}{-m'}\langle F'\|d\|F\rangle.
	\end{multline}
Some people use a different convention for the reduced matrix element which is related to the convention used here by $(F\|d\|F') = \sqrt{2F+1}\langle F\|d\|F'\rangle$. We have here used the following notation for Clebsch-Gordan coefficients $C_{jm\,j'm'}^{JM} = \langle jm\,j'm'|JM\rangle$. It is convenient to define new operators $\boldsymbol\sigma_{F'F}^{+} = \mathbf{d}_{F'F}^{(+)}/\langle F'\|d\|F\rangle$. The adjoint is defined as $\boldsymbol\sigma_{FF'}^{-} = (\boldsymbol\sigma_{F'F}^{+})^{\dag}$. Then the polarizability tensor operator can be expressed as ${\boldsymbol\alpha} = -\sum_{F'\tilde F F}(\langle \tilde F\|d\|F'\rangle\langle F'\|d\|F\rangle/\Delta_{F'F})\boldsymbol\sigma_{\smash{\tilde FF'}}^{-}\!\otimes\!\boldsymbol\sigma_{\smash{F'F}}^{+}$. Using the above definition one obtains 
	\begin{multline}
	\boldsymbol\sigma_{\smash{\tilde FF'}}^{-}\!\otimes\boldsymbol\sigma_{\smash{F'F}}^{+} = (2F'+1)\sum_{p,q}\sum_{n,m,\smash{m'}}(-1)^{F+\tilde F-2+2m'}\\
	\times\tj{\tilde F}{1}{F'}{n}{p}{-m'}\tj{F}{1}{F'}{m}{q}{-m'} |\tilde Fn\rangle\langle Fm|\mathbf{e}_{p}^{\vphantom{*}}\!\otimes\mathbf{e}_{q}^{*}.
	\end{multline}
This expression can now be split into its operator and tensor part by inserting the identity \cite{Hammerer:thesis}
	\begin{equation}
	\sum_{k=0}^{2}\sum_{l=-k}^{k}(2k+1)\tj{1}{1}{k}{-q}{p}{l}\tj{1}{1}{k}{-\tilde q}{\tilde p}{l} = \delta_{p\tilde p}\delta_{q\tilde q}.
	\end{equation}
such that
\begin{widetext}
	\begin{multline}
	\boldsymbol\sigma_{\smash{\tilde FF'}}^{-}\!\otimes\boldsymbol\sigma_{\smash{F'F}}^{+} = (2F'+1)(-1)^{F+F'}\sum_{k,l}(2k+1)\left[\sum_{\tilde p,\tilde q}(-1)^{\tilde q}\tj{1}{1}{k}{-\tilde q}{\tilde p}{l}\mathbf{e}_{\tilde p}^{\vphantom{*}}\otimes\mathbf{e}_{\tilde q}^{*}\right]\\
	\times\left[\sum_{n,m}(-1)^{2\tilde F+F-n}\sum_{p,q,\smash{m'}}(-1)^{F'+2-p-m'-q}
	\tj{F'}{1}{F}{-m'}{q}{m} \tj{1}{1}{k}{-q}{p}{l} \tj{1}{F'}{\tilde F}{-p}{m'}{-n} |\tilde Fn\rangle\langle Fm|\right].
	\end{multline}
The expression for the operator part can be further simplified by evaluating the sum over $p$, $q$, and $m'$ using the identity
	\begin{equation}
	\sum_{\mu_{1},\mu_{2},\mu_{3}}\!\!\!(-1)^{l_{1}+l_{2}+l_{3}-\mu_{1}-\mu_{2}-\mu_{3}}
	\tj{l_{2}}{l_{3}}{j_{1}}{-\mu_{2}}{\mu_{3}}{m_{1}}
	\tj{l_{3}}{l_{1}}{j_{2}}{-\mu_{3}}{\mu_{1}}{m_{2}}
	\tj{l_{1}}{l_{2}}{j_{3}}{-\mu_{1}}{\mu_{2}}{m_{3}}
	=
	\tj{j_{1}}{j_{2}}{j_{3}}{m_{1}}{m_{2}}{m_{3}}
	\Gj{j_{1}}{j_{2}}{j_{3}}{l_{1}}{l_{2}}{l_{3}}.
	\end{equation}
Changing 3$j$-symbols to Clebsh-Gordan coefficients we get
	\begin{equation}
	\boldsymbol\sigma_{\smash{\tilde FF'}}^{-}\otimes\boldsymbol\sigma_{\smash{F'F}}^{+} = (2F'+1)\frac{(-1)^{F+F'+1}}{\sqrt{3(2\smash{\tilde F} + 1)\vphantom{\big[}}}\sum_{k=1}^{2}(2k+1)
	\Gj{F}{k}{\tilde F}{1}{F'}{1}
	\sum_{l=-k}^{k}\left[\sum_{n,m}C_{Fm\,kl}^{\tilde Fn}\, |\tilde Fn\rangle\langle Fm|\right]\left[\sum_{p,q}C_{1p\,kl}^{1q}\, \mathbf{e}_{ p}^{\vphantom{*}}\otimes\mathbf{e}_{ q}^{*}\right].
	\end{equation}
Finally, the overall polarizability can be written as $\displaystyle{\boldsymbol\alpha} = -(2J'+1)|\langle J'\|d\|J\rangle|^{2}\frac{1}{\Delta}\sum_{k,\smash{\tilde F}}a_{k}^{F\tilde F}(\Delta)\,\mathbf T^{k}_{F\tilde F}$, where
	\begin{equation}
	\label{a_coefficients_all}
	{a}_{k}^{F\tilde F}(\Delta) = -(-1)^{F}(2k+1)c_{k}\sqrt{\frac{2F+1}3}\sum_{F'}\frac{(-1)^{F'}(2F'+1)}{1-\nicefrac{\delta_{F'}}{\Delta}}
	\times\Gj{J'}{F'}{I}{F}{J}{1} \Gj{J'}{F'}{I}{\tilde F}{J}{1} \Gj{F}{k}{\tilde F}{1}{F'}{1}.
	\end{equation}
\end{widetext}
The hyperfine splitting is $\delta_{F'} = \Delta - \Delta_{F'}$. 
Here we have used that
	\begin{multline}
	\langle \tilde F\|d^{(+)}\|F'\rangle\langle F'\|d^{(-)}\|F\rangle = \\
	=(-1)^{2(J'+I+F)+\tilde F-F'}
	\Gj{J'}{F'}{I}{F}{J}{1} \Gj{J'}{F'}{I}{\tilde F}{J}{1}\quad\\
	 \times \sqrt{(2F+1)(2\smash{\tilde F}+1)\vphantom{\big|}}\,(2J'+1)\big|\langle  J'\|d\|J\rangle\big|^{2}.
	\end{multline}
The coefficients $c_{k}$ read
	\begin{equation}
	\label{app:coeff_ck}
	\begin{split}
	c_{0} &= 1,\qquad
	c_{1} = \frac1{\sqrt{2F(F+1)}},\\
	c_{2} &= \frac3{\sqrt{10F(F+1)(2F-1)(2F+3)}}.
	\end{split}
	\end{equation}
These coefficients are chosen in such a way that the resulting Hamiltonian contains spin component as irreducible tensors. The irreducible tensors are defined as follows
	\begin{equation}
	\label{tensor_irr}
	\mathbf T^{k}_{F\tilde F} = \frac1{c_{k}}\sum_{l=-k}^{k}M^{k}_{l;F\tilde F}\sum_{p,q} C_{1p\,kl}^{1q}\,\mathbf{e}_{p}^{\vphantom{*}}\!\otimes\mathbf{e}_{q}^{*}
	\end{equation}
For $\tilde F=F$ we have
	\begin{equation*}
	M^{k}_{l;FF} = \sum_{n,m}C_{Fm\,kl}^{Fn}\, |Fn\rangle\langle Fm|
	\end{equation*}
if $\tilde F\neq F$
	\begin{multline}
	 M^{k}_{l;F\tilde F} = \sum_{n,m}\left(C_{Fm\,kl}^{\tilde F n}\, |\tilde Fn\rangle\langle Fm|\right.\\
	+ \left.(-1)^l C_{Fm;k,-l}^{\tilde Fn}\, |Fm\rangle\langle \tilde Fn|\right)
	\end{multline}

\section{Transformation of the light equation of motion}
\label{A:light-transform}
We will work
in the paraxial approximation assuming a flat transverse profile and write the forward electric field
as
	\begin{align}
	\mathbf{E}_{\rm{F}}(z,t) &= \sqrt{\frac{\omega_0}{2\epsilon_0A}}\sum_{\sigma}\!\int\!\frac{d^{3}k\,d^{2}\rho}{(2\pi)^{3}}\boldsymbol{\epsilon}_{\sigma} (a_{k\sigma}(t)e^{i(k_{z}z+k_{\!\perp}\rho)}+h.c.),\nonumber\\
	&=|E|\sum_{\sigma}\boldsymbol{\epsilon}_{\sigma}(a_{\sigma}(z,t)+a^{\dagger}_{\sigma}(z,t)).
	\end{align}
Here $|E|=\sqrt{\frac{\omega_0}{2\epsilon_0A}}$ and we have defined the space dependent operators
$a_{\sigma}(z,t)=\int\frac{dk_{z}}{2\pi}a_{k_{z}\sigma}(t)e^{ik_{z}z}$, where it
is assumed that the different $k$ are close to $k_0$, so that the operators
oscillate at the common frequency $\omega_0=|k_0|$. For the radiation
field we have that
\begin{equation}
H_{\rm{L}}=\sum_{\sigma}\int\frac{dk}{2\pi}\omega_ka_{k\sigma}^{\dagger}a_{k\sigma}^{\phantom{\dagger}}.
\end{equation}
Now we are to ready to form the EOM
\begin{align}
\frac{\partial}{\partial{t}}{a_{\sigma}(z,t)}&=i[H_{\rm{int}}+H_{\rm{L}},a_{\sigma}(z,t)],\\
{}[H_{\rm{L}},a_{\sigma}(z,t)]&=\int\frac{dk}{2\pi}[H_{\rm{L}},a_{k\sigma}(t)]e^{ikz}\nonumber\\&=-\int\frac{dk}{2\pi}\omega_ka_{k\sigma}(t)e^{ikz}.
\end{align}
But we also have from the explicit $z$-dependence that
\begin{equation}
\frac{\partial}{\partial{z}}{a}_{\sigma}(z,t)=\int\frac{dk}{2\pi}ika_{k\sigma}(t)e^{ikz},
\end{equation}
which we recognize as $-i[H_{\rm{L}},{a}_{\sigma}(z,t)]$. So we replace the time
evolution from the radiation field by minus the derivative with respect to
$z$ and end up with
\begin{equation}
\left(\frac{\partial}{\partial{t}}+\frac{\partial}{\partial{z}}\right){a}_{\sigma}(z,t)=i[H_{\rm{int}},{a}_{\sigma}(z,t)].
\end{equation}
This procedure is directly applicable for Stokes operators too. In the end we throw away the time derivative, which is the same as ignoring
retardation effects, but this approimation can even be made exact by introducing a suitable rescaled time. By doing this we have thus an equation in 
position and not time for light observables.

\section{Calculations of atomic spin noise and decay rate coefficients for light and atoms}
\label{app:alpha}
\subsection{Light matrix $\alpha^{2}$}

The dimensionless tensor polarizability can be written as 
	\begin{equation}\label{app:alpha_dimensionless}
	{\alpha} = \sum_{k,\smash{\tilde F}}a_{k}^{F\tilde F}(\Delta)\,\mathbf T^{k}_{F\tilde F}.
	\end{equation}
We assume that the atomic spin is polarized along $x$ axis and during interaction the collective spin experiences merely small rotations. Having the quantization axis to be the $x$-axis we have to linearize the spin dependent operators around the initial state $|FF\rangle$. Using the expression for the irreduccible tensors \eqref{tensor_irr}  and averaging over the initial spin state we obtain
	\begin{multline}
	\langle{\alpha}^{2}\rangle = \sum_{k,k',\smash{\tilde F}} \frac{a_{k}^{F\tilde F}a_{k'}^{F\tilde F}}{c_{k}c_{k'}}\sum_{p,q}\sum_{\smash{r,s,l,l'}}
	C_{\tilde F r\,kl}^{FF}C_{FF\,k'l'\vphantom{\tilde F}}^{\tilde F r}\\
	\times C_{\smash{1p\,kl}}^{1s}C_{\smash{1s\,k'l'}}^{1q}
	\mathbf{e}_{p}^{\vphantom{*}}\!\otimes\mathbf{e}_{q}^{*}.
	\end{multline}
We have changed the notation for Clebsch-Gordan coefficients to the shorter form $\langle jm\,j'm'|JM\rangle = C_{jm\,j'm'}^{JM}$. Taking into account restrictions on momentum projections given by Clebsch-Gordans the sum over four indexes is reduced to a sum over a single index. We have $p=q$, $l=s-q$, $l'=-l$,  $r=F+q-s$, and hence
	\begin{multline}
	\label{app:alpha_spherical}
	\langle{\alpha}^{2}\rangle = \sum_{k,k',\smash{\tilde F}} \frac{a_{k}^{F\tilde F}a_{k'}^{F\tilde F}}{c_{k}c_{k'}}\sum_{q,s}
	C_{\tilde F, F+q-s;k,s-q}^{FF}C_{FF;k'\!,q-s\vphantom{\tilde F}}^{\tilde F,F+q-s}\\
	\times C_{\smash{1q;k,s-q}}^{1s}C_{\smash{1s;k'\!,q-s}}^{1q}
	\mathbf{e}_{q}^{\vphantom{*}}\!\otimes\mathbf{e}_{q}^{*}.
	\end{multline}
The required elements of the matrix $\alpha^{2}$ for the atomic spin aligned in the $x$-direction in Cartesian coordinates read
	\begin{align}
	\label{app:alpha_dekartX}
	\langle\alpha^{2}\rangle_{xx} &= \langle\alpha^{2}\rangle_{00}\\
	\label{app:alpha_dekartY}
	\langle\alpha^{2}\rangle_{yy} &= \frac12\left(\langle\alpha^{2}\rangle_{--}	+ \langle\alpha^{2}\rangle_{++}\right)\\
	\langle\alpha^{2}\rangle_{xy} &= 0.
	\end{align}
	
\subsection{Atomic Langevin noise $\left(i[\alpha,j_{i}]\right)^{2}$}\label{app:atomic_noise}

Let us consider the matrix $\zeta_{i}=i[\alpha,j_{i}]$. The commutator of an irreducible tensor and spherical spin components reads
	\begin{equation}
	[M^{k}_{l;FF},j_{\mu}] = -\sqrt{k(k+1)}C^{k,l+\mu}_{kl;1\mu}\,M^{k}_{l+\mu;FF}.
	\end{equation}
This is true for an irreducible tensor defined for a single spin state therefore we consider the case of $\tilde F=F$.
	\begin{multline}
	\label{g-matrix}
	\boldsymbol \zeta_{\mu}^{FF} = -i\sum_{k}\frac{a_{k}^{FF}}{c_{k}}\sqrt{k(k+1)}\sum_{l,p,q}\\
	\times C^{k,l+\mu}_{kl;1\mu}\,M^{k}_{l+\mu;FF} C_{1p;kl}^{1q}\mathbf{e}_{p}^{\vphantom{*}}\!\otimes\mathbf{e}_{q}^{*}.
	\end{multline}
Using the same method as in the previous subsection one obtains
	\begin{align}
	\langle\boldsymbol \zeta_{\mu}\boldsymbol \zeta_{\nu}\rangle^{FF} &= -\sum_{k,k'} \frac{a_{k}^{F F}a_{k'}^{F F}}{c_{k}c_{k'}}\sqrt{k(k+1)k'(k'+1)}\sum_{p,q}\sum_{\smash{r,s,l,l'}}\nonumber\\
	&\phantom{=}\times
	C_{kl;1\mu}^{k,l+\mu}C_{k'l';1\nu}^{k'\!\!,l'+\nu}
	C_{F r;k,l+\mu}^{FF}C_{FF;k'l'+\nu}^{F r}\nonumber\\
	&\phantom{=}\times
	C_{\smash{1p;kl}}^{1s}C_{\smash{1s;k'l'}}^{1q}
	\mathbf{e}_{p}^{\vphantom{*}}\!\otimes\mathbf{e}_{q}^{*}
	\end{align}
	After some transformations one comes to the following expression
	\begin{align}
	\label{app:langevin_spherical1}
	\langle \zeta_{\mu}\zeta_{\nu}\rangle_{pq}^{FF} &= -\sum_{k,k'} \frac{a_{k}^{F F}a_{k'}^{F F}}{c_{k}c_{k'}}\sqrt{k(k+1)k'(k'+1)}\sum_{s}\nonumber\\
	&\phantom{=}\times \delta_{p,q+\mu+\nu}
	C_{k,s-p;1\mu}^{k,s-q-\nu}C_{k'\!,q-s;1\nu}^{k'\!\!,q-s+\nu}\nonumber\\
	&\phantom{=}\times
	C_{F,F+q-s+\nu;k,s-q-\nu}^{FF}C_{FF;k'\!,q-s+\nu}^{F,F+q-s+\nu}\nonumber\\
	&\phantom{=}\times
	C_{\smash{1,p;k,s-p}}^{1s}C_{\smash{1s;k'\!,q-s}}^{1q}
	\end{align}
The case of $\tilde F\neq F$ has to be considered specifically. The commutator with spin components doesn't have a nice form anymore
	\begin{multline*}
	[M^{k}_{l;F\smash{\tilde F}},j_{\mu}] = \sqrt{F(F+1)}\sum_{m,n,r}\left[ C_{Fr;kl}^{\tilde Fn}C_{Fm;1\mu}^{ Fr}|\tilde Fn\rangle\langle Fm|\right.\\
	- \left.(-1)^{l}C_{Fr;k-l}^{\tilde Fn} C_{Fr;1\mu}^{ Fm}|Fm\rangle\langle \tilde Fn|\right].
	\end{multline*}
Using this expression for the commutator and doing some algebra we obtain
	\begin{align}
	\label{app:langevin_spherical2}
	\langle \zeta_{\mu}\zeta_{\nu}\rangle_{pq}^{F\tilde F} &= (-1)^{s-p}\delta_{p,q+\mu+\nu}F(F+1) \nonumber\\
	&\phantom{=}\times
	C_{F,F-\mu;1\mu}^{FF}C_{FF;1\nu}^{F,F-\nu}
	\sum_{k,k'}\frac{a_{k}^{F \tilde F}a_{k'}^{F\tilde F}}{c_{k}c_{k'}}\nonumber\\
	&\phantom{=}\times\sum_{s}
	C_{F,F-\mu;k,p-s}^{\tilde F,F+q-s+\nu} C_{F,F+\nu;k'\!,q-s}^{\tilde F,F+q-s+\nu}\nonumber\\
	&\phantom{=}\times
	C_{\smash{1,p;k,s-p}}^{1s}C_{\smash{1s;k'\!,q-s}}^{1q}.
	\end{align}
The full matrix $\langle \zeta^{2}\rangle$ is just a sum of all of these contributions for different $\tilde F$.
	\begin{equation}
	\label{app:langevin_spherical3}
	\langle \zeta_{\mu}\zeta_{\nu}\rangle_{pq} = \sum_{\tilde F}\langle \zeta_{\mu}\zeta_{\nu}\rangle_{pq}^{\smash{F\tilde F}}.
	\end{equation}
Since the spin is prepared in a polarized state along the $x$ direction the spin noise coefficient for $i$-th spin component in presence of the driving field polarized parallel to the spin is given by the $xx$ components of the matrix  $\langle \zeta^{2}_{i}\rangle$. The case of the orthogonal configuration of the spin and light is described by the $yy$ element of the matrix.
	\begin{align}
	\label{app:langevin_dekart1}
	\langle \zeta_{x}^{2}\rangle_{\parallel}  &= \langle \zeta_{0}\zeta_{0}\rangle_{00},\\
	\langle \zeta_{y}^{2}\rangle_{\parallel} &= -\frac12\left\{\langle \zeta_{-}\zeta_{+}\rangle_{00} + \langle \zeta_{+}\zeta_{-}\rangle_{00}\right\},\\
	\langle \zeta_{z}^{2}\rangle_{\parallel} &= \langle \zeta_{y}^{2}\rangle_{\parallel},\\
	\langle \zeta_{x}^{2}\rangle_{\perp} & = \frac12\left\{\langle \zeta_{0}\zeta_{0}\rangle_{--} + \langle \zeta_{0}\zeta_{0}\rangle_{++}\right\},\\
	\langle \zeta_{y}^{2}\rangle_{\perp} &= -\frac14\left\{\langle \zeta_{-}\zeta_{-}\rangle_{-+} + \langle \zeta_{+}\zeta_{+}\rangle_{+-} + \langle \zeta_{-}\zeta_{+}\rangle_{--}\right.\nonumber\\
	 &\phantom{=}+ \left.\langle \zeta_{-}\zeta_{+}\rangle_{++} + \langle \zeta_{+}\zeta_{-}\rangle_{--} + \langle \zeta_{+}\zeta_{-}\rangle_{++}\right\},\\
	\langle \zeta_{z}^{2}\rangle_{\perp} &= \frac14\left\{\langle \zeta_{-}\zeta_{-}\rangle_{-+} + \langle \zeta_{+}\zeta_{+}\rangle_{+-} - \langle \zeta_{-}\zeta_{+}\rangle_{--}\right.\nonumber\\
	\label{app:langevin_dekart2}
	 &\phantom{=}- \left.\langle \zeta_{-}\zeta_{+}\rangle_{++} - \langle \zeta_{+}\zeta_{-}\rangle_{--} - \langle \zeta_{+}\zeta_{-}\rangle_{++}\right\}.
	\end{align}

\subsection{Spin decay rates and $\xi_{i}$}\label{app:xi}
In this subsection we derive expressions for the spin decay rates. The coefficients for the spin components decays are defined via linearization of the operator $\xi_{i} = \alpha^{2}j_{i}+j_{i}\alpha^{2}-2\alpha j_{i}\alpha = -i[\alpha,\zeta_{i}]$ for the given polarized spin state. We want to have $\xi_{i}\sim\Xi_{i}j_{i}$ which implies
	\begin{equation}
	\langle\xi_{i}\rangle =  \Xi_{i}\langle j_{i}\rangle,\qquad
	\langle\xi_{i}j_{i}\rangle =  \Xi_{i}\langle j_{i}^{2}\rangle.
	\end{equation}
The last equality is always nontrivial so we can use it for defining $\Xi_{i} = {\langle\xi_{i}j_{i}\rangle}/{\langle j_{i}^{2}\rangle}$. Using expressions \eqref{app:alpha_dimensionless} and \eqref{g-matrix} we proceed
	\begin{align}
	\boldsymbol\xi_{\mu}^{FF} &= -\sum_{k,k'} \frac{a_{k}^{F F}a_{k'}^{F F}}{c_{k}c_{k'}}\sqrt{k(k+1)}\sum_{p,q}\sum_{\smash{r,s,l,l'}} C_{kl;1\mu}^{k,l+\mu}\nonumber\\
	&\phantom{=}\times\left[
	C_{Fr;k'l'}^{Fn} C_{Fm;k,l+\mu}^{Fr} C_{1p;k'l'}^{1s} C_{1s;kl}^{1q} \right.\nonumber\\
	&\phantom{=} - \left.C_{Fr;k,l+\mu}^{Fn} C_{Fm;k'l'}^{Fr} C_{1p;kl}^{1s} C_{1s;k'l'}^{1q}\right]\nonumber\\
	&\phantom{=}\times |Fn\rangle\langle Fm|\,
	\mathbf{e}_{p}^{\vphantom{*}}\!\otimes\mathbf{e}_{q}^{*}.
	\end{align}
After averaging over the initial spin polarized state $|FF\rangle$ we obtain the required matrix elements in the spherical basis
	\begin{align}
	\langle\xi_{\mu}j_{\nu}\rangle_{pq}^{FF} &=
	-\sum_{k,k'} \frac{a_{k}^{F F}a_{k'}^{F F}}{c_{k}c_{k'}}\sqrt{k(k+1)F(F+1)}\nonumber\\
	&\phantom{=}\times \sum_{r,s,l,l'} C_{kl;1\mu}^{k,l+\mu} C_{FF;1\nu}^{F,F+\nu}\nonumber\\
	&\phantom{=}\times\left[
	C_{Fr;k'l'}^{FF} C_{F,F+\nu;k,l+\mu}^{Fr} C_{1p;k'l'}^{1s} C_{1s;kl}^{1q} \right.\nonumber\\
	 &\phantom{=}- \left.C_{Fr;k,l+\mu}^{FF} C_{F,F+\nu;k'l'}^{Fr} C_{1p;kl}^{1s} C_{1s;k'l'}^{1q}\right].
	\end{align}
The Clebsch-Gordans require the following constraints on the indexes to be fulfilled for the first term $r=F+p-s,\,l=q-s,\,l'=s-p$ and for the second term $r=F+q-s+\nu,\,l=s-p,\,l'=q-s$ plus the usual condition $p=q+\mu+\nu$. Finally, we obtain
	\begin{align}
	\label{app:decay_spin_spherical1}
	\langle\xi_{\mu}j_{\nu}\rangle_{pq}^{FF} &=
	-\delta_{p,q+\mu+\nu}
	\sum_{k,k'} \frac{a_{k}^{F F}a_{k'}^{F F}}{c_{k}c_{k'}} C_{FF;1\nu}^{F,F+\nu}\nonumber\\
	&\phantom{=}\times \sqrt{k(k+1)F(F+1)} \sum_{s}  \nonumber\\
	&\phantom{=}\times\left[
	C_{F,F+p-s;k',s-p}^{FF} C_{F,F+\nu;k,q-s+\mu}^{F,F+p-s} \right.\nonumber\\
	&\phantom{=}\times C_{k,q-s;1\mu}^{k,q-s+\mu} C_{1p;k',s-p}^{1s} C_{1s;k,q-s}^{1q} \nonumber\\
	 &\phantom{=}- C_{F,F+q-s+\nu;k,s-p+\mu}^{FF} C_{F,F+\nu;k',q-s}^{F,F+q-s+\nu}\nonumber\\
	 &\phantom{=}\times\left. C_{k,s-p;1\mu}^{k,s-p+\mu}  C_{1p;k,s-p}^{1s} C_{1s;k',q-s}^{1q}\right].
	\end{align}
In case of $\tilde F\neq F$ essentially the same calculations provide us with the necessary expression
	\begin{align}
	\label{app:decay_spin_spherical2}
	\langle\xi_{\mu}j_{\nu}\rangle_{pq}^{F\tilde F} &=
	\delta_{p,q+\mu+\nu}
	\sum_{k,k'} \frac{a_{k}^{F \tilde F}a_{k'}^{F \tilde F}}{c_{k}c_{k'}}{F(F+1)} C_{FF;1\nu}^{F,F+\nu}\nonumber\\
	&\phantom{=}\times \sum_{s}(-1)^{s-p}C_{1p;k,s-p}^{1s} C_{1s;k',q-s}^{1q}  \nonumber\\
	&\phantom{=}\times\left[
	C_{FF;k,p-s}^{\tilde F,F-s+p} C_{F,F+\mu+\nu;k',q-s}^{\tilde F,F-s+p} C_{F,F+\nu;1\mu}^{F,F+\mu+\nu}  \right. \nonumber\\
	 &\phantom{=}+ \left. C_{F,F-\mu;1\mu}^{FF} C_{F,F-\mu;k,p-s}^{\tilde F,F+q-s+\nu} C_{F,F+\nu;k',q-s}^{\tilde F,F+q-s+\nu}\right].
	\end{align}
The full matrix $\langle\xi_{\mu}j_{\nu}\rangle$ can be found by summation over all $\tilde F$
	\begin{equation}
	\label{app:decay_spin_spherical3}
	\langle\xi_{\mu}j_{\nu}\rangle_{pq} = \sum_{\tilde F}\langle\xi_{\mu}j_{\nu}\rangle_{pq}^{\smash{F\tilde F}}.
	\end{equation}
We recall that the spin is prepared in a polarized state along the $x$ direction. Then the decay rate coefficient for $j_{i}$ in presence of the driving field polarized parallel to the spin is given by the $xx$-component of the matrix  $\langle\xi_{i}j_{i}\rangle$. The case of the orthogonal configuration of the spin and light is described by $yy$-element of the matrix.
	\begin{align}
	\label{app:decay_spin_dekart1}
	\Xi_{x_{\parallel}} &= \frac1{F^{2}}{\langle \xi_{0}j_{0}\rangle_{00}},\\
	\Xi_{y_{\parallel}} &= -\frac1F\left\{\langle \xi_{-}j_{+}\rangle_{00} + \langle \xi_{+}j_{-}\rangle_{00}\right\},\\
	\Xi_{z_{\parallel}} &= \Xi_{y_{\parallel}},\\
	\Xi_{x_{\perp}}  &= \frac1{2F^{2}}\left\{\langle \xi_{0}j_{0}\rangle_{--} + \langle \xi_{0}j_{0}\rangle_{++}\right\},\\
	\Xi_{y_{\perp}} &= -\frac1{2F}\left\{\langle \xi_{-}j_{-}\rangle_{-+} + \langle \xi_{+}j_{+}\rangle_{+-} + \langle \xi_{-}j_{+}\rangle_{--}\right.\nonumber\\
	&\phantom{=} + \left.\langle \xi_{-}j_{+}\rangle_{++} + \langle \xi_{+}j_{-}\rangle_{--} + \langle \xi_{+}j_{-}\rangle_{++}\right\},\\
	\label{app:decay_spin_dekart2}
	\Xi_{z_{\perp}} &= \frac1{2F}\left\{\langle \xi_{-}j_{-}\rangle_{-+} + \langle \xi_{+}j_{+}\rangle_{+-} - \langle \xi_{-}j_{+}\rangle_{--}\right.\nonumber\\
	 &\phantom{=}- \left.\langle \xi_{-}j_{+}\rangle_{++} - \langle \xi_{+}j_{-}\rangle_{--} - \langle \xi_{+}j_{-}\rangle_{++}\right\}.
	\end{align}

\subsection{Relation between the spin decay and the spin noise}
\label{app:spin_deacy_to_noise_relation}
Let us consider the commutator of the Langevin noises for spin polarized atoms $\langle[F_{i},F_{j}]\rangle$. The noise is defined by the matrix $\zeta_{i}=i[\alpha,j_{i}]$ calculated in Appendix~\ref{app:atomic_noise}. Therefore, one can consider the mean value of the commutator $\langle[\zeta_{y},\zeta_{z}]\rangle = -\langle[[\alpha,j_{y}],[\alpha,j_{z}]]\rangle$. On the over hand, the spin decay rate is given by the operator $\xi_{i} = \alpha^{2}j_{i}+j_{i}\alpha^{2}-2\alpha j_{i}\alpha$. In the previous subsection of the appendix we have shown that for spin polarized atoms the operator is proportional to the spin projection $\langle\xi_{i}\rangle\sim\Xi_{i}\langle j_{i}\rangle$. Using this fact one can show that
	\begin{equation}
	\langle[\zeta_{y},\zeta_{z}]\rangle = \frac{i}2(\Xi_{y} + \Xi_{z} - \Xi_{x})\langle j_{x}\rangle.
	\end{equation}
This holds true for any cyclic permutation of $x,\,y,\,z$.

\section{Expressions for $^{133}$Cs}\label{app:explicit_expression}
In this section we evaluate the $A,\,B,\,C$ coefficients for $^{133}$Cs optically pumped to the ground state $F=4$. In this case we have $I=7/2$, $J=1/2$, $J'=3/2$, $F=4$, $\tilde F=3$. In order to simplify the expressions we use the following notations: $a_{k}$ for $a_{k}^{FF}$ and $b_{k}$ for $a_{k}^{F\tilde F}$.
	\begin{align}
	a_{0} &= \frac7{144}\left(\frac{44}{21} + \frac{1}{1-\frac{\Delta _{45}}{\Delta }}+\frac{1}{3 \left(1-\frac{\Delta_{35}}{\Delta }\right)}\right),\\
	a_{1} &= \frac{7}{5760}\left(\frac{176}7 - \frac{3}{1-\frac{\Delta_{45}}{\Delta}} - \frac{5}{1-\frac{\Delta_{35}}{\Delta}}\right),\\
	a_{2} &= \frac1{5760}\left(16-\frac{21}{1-\frac{\Delta _{45}}{\Delta }}+\frac{5}{1-\frac{\Delta_{35}}{\Delta }}\right),
	\end{align}
	\begin{align}
	b_{1} &= \frac1{128\sqrt5}\left(\frac{5}{1-\frac{\Delta_{45}}{\Delta }}+\frac{3}{1-\frac{\Delta_{35}}{\Delta }}\right),\\
	b_{2} &= \frac3{128\sqrt{77}}\left(\frac{1}{1-\frac{\Delta _{45}}{\Delta }}-\frac{1}{1-\frac{\Delta
   _{35}}{\Delta }}\right).
	\end{align}
Light coefficients $A_{\mu}$:
	\begin{align}
	 A_{x} &= \frac1{4a_{1}^{2}}[a_0^2+ 4 a_1^2 + 56 a_1 a_2 - \frac{112}{3}a_0 a_2 +\frac{4900}{9} a_2^2\nonumber\\
	&\phantom{=}+ \frac{140}9(b_1^2 +\frac{77 }{5} b_2^2 + 2 \sqrt{\frac{77}{5}} b_1 b_2)],\\
	 A_{y} &= \frac1{4a_{1}^{2}}[a_0^2 +18 a_1^2 -28 a_1 a_2 +\frac{56}{3} a_{0}a_2 + \frac{2170}{9} a_2^2\nonumber\\
	&\phantom{=}+ \frac{70}9(b_1^2+\frac{539 }{15}b_2^2 - 2\sqrt{\frac{77}{5}} b_1 b_2)].
	\end{align}
	
Spin decay coefficients $B_{i}$:
	\begin{align}
	B_{x_{\parallel}} &= \frac1{2a_{1}^{2}}[a_1^2+14 a_2 a_1+49 a_2^2\nonumber\\
	&\phantom{=}+\frac{140}9 (b_1^2+\frac{77 }{5}b_2^2 + 2 \sqrt{\frac{77}{5}} b_1 b_2)],\\
	B_{y_{\parallel}} &= \frac1{4a_{1}^{2}}[a_1^2-98 a_2 a_1+273 a_2^2\nonumber\\
	&\phantom{=}+\frac{245}9(b_1^2+\frac{55 }{3}b_2^2+2 \sqrt{\frac{55}{7}} b_1 b_2)],\\
	B_{z_{\parallel}} &= B_{y_{\parallel}},\\
	 B_{x_{\perp}} &= \frac1{4a_{1}^{2}}[a_1^2-14 a_2 a_1+105 a_2^2\nonumber\\
	&\phantom{=}+\frac{140}9 (b_1^2+\frac{539 }{15}b_2^2-2 \sqrt{\frac{77}{5}} b_1 b_2)],\\
	 B_{y_{\perp}} &= \frac1{2a_{1}^{2}}[ a_1^2+ 56 a_2 a_1-35 a_2^2\nonumber\\
	&\phantom{=}+\frac{175}{18}(b_1^2+\frac{693 }{25}b_2^2-\frac{2}{5} \sqrt{\frac{77}{5}}b_1 b_2)],\\
	 B_{z_{\perp}} &= B_{x_{\perp}}.
	\end{align}
	
Spin noise coefficients $C_{i}$:
	\begin{align}
	 C_{x_{\parallel}} &= \frac1{a_{1}^{2}}[a_1^2+14 a_2 a_1 + 49a_2^2\nonumber\\
	 &\phantom{=}+ \frac{560}9 (b_1^2+\frac{77 }{5}b_2^2+2 \sqrt{\frac{77}{5}} b_1 b_2)],\\
	 C_{y_{\parallel}} &= \frac1{a_{1}^{2}}[4 a_1^2+308 a_2^2\nonumber\\
	&\phantom{=}+\frac{35}6( b_1^2+\frac{1001}{45}b_2^2 + \frac{2}{3} \sqrt{\frac{77}{5}} b_1 b_2)],\\
	 C_{z_{\parallel}} &= C_{y_{\parallel}},
	\end{align}\\	
\begin{widetext}
	\begin{align}
	 C_{x_{\perp}} &= \frac1{2a_{1}^{2}}[a_1^2-14 a_2 a_1+161 a_2^2
	+\frac{560}9 (b_1^2+\frac{539}{15}b_2^2 -2 \sqrt{\frac{77}{5}} b_1 b_2)],\\
	 C_{y_{\perp}} &= \frac1{2a_{1}^{2}}[9 a_1^2-14 a_2 a_1+63 a_2^2
	+ \frac{175}{18}(b_1^2+\frac{693 }{25}b_2^2 -\frac{2}{5} \sqrt{\frac{77}{5}}b_1 b_2)],\\
	 C_{z_{\perp}} &= \frac1{2a_{1}^{2}}[a_1^2+14 a_2 a_1+651 a_2^2
	+ \frac{175}{18}(b_1^2+\frac{693 }{25}b_2^2 -\frac{2}{5} \sqrt{\frac{77}{5}}b_1 b_2)].
	\end{align}
\end{widetext}
	
\bibliographystyle{unsrt}
\bibliography{references} 

\end{document}